\documentclass[onecolumn]{elsart3p}

\usepackage{graphicx,amssymb}
\usepackage{amssymb}
\newcommand{\om}{\omega}
\newcommand{\D}{\mathrm{d}}
\newcommand{\I}{\mathrm{i}}

\DeclareMathSymbol{\varPhi}{\mathalpha}{operators}{"08}
\DeclareMathSymbol{\varOmega}{\mathalpha}{operators}{"0A}

\makeatletter
\def\myfootnotesize{\@setsize\footnotesize{10pt}\ixpt\@ixpt} % 9 point on 10

\@ifundefined{square}{}{\let\Box\square}
\makeatother

\newcommand{\IJMP}[1]{{Int. J. Mod. Phys. A \textbf{#1}}}

\newcommand{\npb}[1]{{Nucl. Phys. B \textbf{#1}}}

\newcommand{\SJNP}[1]{{Sov. J. Nucl. Phys. \textbf{#1}}}

\def\amu{a_\mu}
\def\amuh{a_\mu^{{\mathrm{had}}}}

\def\dah{\Delta\alpha^{(5)}_{\rm had}}

\def\dah0{\Delta\alpha^{(5)}_{\rm had}(-s_0)}
\def\sf{spectral function}

\newcommand{\mv}{\mbox{MeV}}

\newcommand{\MSb}{$\overline{\mathrm{MS}}$ }

\newcommand{\epm}{e^+e^-}
\newcommand{\pipi}{\pi^+\pi^-}

\newcommand{\eepp}{e^+e^- \rightarrow \pi^+\pi^-}
\newcommand{\be}{\begin{equation}}
\newcommand{\ee}{\end{equation}}
\newcommand{\ba}{\begin{eqnarray}}
\newcommand{\ea}{\end{eqnarray}}
\newcommand{\bea}{\begin{eqnarray*}}
\newcommand{\eea}{\end{eqnarray*}}
\newcommand{\bet}{\begin{center} \begin{tabular}}
\newcommand{\ent}{\end{tabular} \end{center}}
\newcommand{\bb}{\begin{thebibliography}}
\newcommand{\eb}{\end{thebibliography}}

\newcommand{\ra}{\rightarrow}

\newcommand{\bit}{\begin{itemize}}
\newcommand{\eit}{\end{itemize}}

\newcommand{\noi}{\noindent}

\newcommand{\crn}{\nonumber \\}

\newcommand{\ha}{\frac{1}{2}}

\newcommand{\bary}{\begin{array}}
\newcommand{\eary}{\end{array}}

\newcommand{\ep}{\;\: .}
\newcommand{\epo}{\;\: .}
\newcommand{\semis}{\;\:;\;\; }
\newcommand{\comas}{\;\:,\;\; }
\newcommand{\cs}{\;\:, }
\newcommand{\Impa}{\mbox{Im} \:}
\newcommand{\Repa}{\mbox{Re} \:}
\renewcommand{\thefootnote}{\fnsymbol{footnote}}
\newcommand{\cB}{{\cal B}}
\newcommand{\cL}{{\cal L}}

\newcommand{\cM}{{\cal M}}
\newcommand{\corr}{\;\;\hat{=}\;\;}
\newcommand{\mr}{M_\rho^2}
\newcommand{\mo}{M_\omega^2}
\newcommand{\power}[1]{\times 10^{#1}}

\newcommand{\Fpiesq}{$|F_{\pi}^{(e)}(s)|^2$}
\newcommand{\Fpitsq}{$|F_{\pi}^{(\tau)}(s)|^2$}

\newcommand{\rogami}{$\rho^0 - \gamma$ mixing }

\journal{}

\begin{document}
%~

\begin{frontmatter}
\begin{flushright}
{\normalsize \rm
DESY 11-008\\
HU-EP-11/04\\
\vspace*{0.5cm}}
\end{flushright}

\title{$\rho^0-\gamma$ mixing in the neutral channel pion form factor $F^{(e)}_\pi(s)$
and its role in comparing $\epm$ with $\tau$ spectral functions}
\author[Berlin,Zeuthen]{Fred Jegerlehner\corauthref{cor}},
\corauth[cor]{Corresponding author.}
\ead{fjeger@physik.hu-berlin.de}
\ead[url]{www-com.physik.hu-berlin.de/\~{}fjeger/}
\author[Katowice]{Robert Szafron}
\ead{rszafron@us.edu.pl}

\address[Berlin]{Humboldt-Universit\"at zu Berlin, Institut f\"ur Physik,
       Newtonstrasse 15, D-12489 Berlin, Germany}
\address[Zeuthen]{Deutsches Elektronen-Synchrotron (DESY), 
Platanenallee 6, D-15738 Zeuthen, Germany}
\address[Katowice]{Institute of Physics, University of Silesia,
ul. Uniwersytecka 4, PL-40007 Katowice, Poland}

\begin{abstract}
We study the effect of $\rho^0-\gamma$ mixing in $e^+e^- \to
\pi^+\pi^-$ and its relevance for the comparison of the square modulus
of the pion from-factor \Fpiesq, as measured in $e^+e^-$ annihilation
experiments, and \Fpitsq~ the corresponding quantity obtained after
accounting for known isospin breaking effects by an isospin rotation
from the $\tau$-decay spectra. After correcting the $\tau$ data for
the missing $\rho-\gamma$ mixing contribution, besides the other known
isospin symmetry violating corrections, the $\pi\pi$ I=1 part of the
hadronic vacuum polarization contribution to the muon $g-2$ are fully
compatible between $\tau$ based and $\epm$ based evaluations. $\tau$
data thus confirm result obtained with $\epm$ data. Our evaluation of
the leading order vacuum polarization contribution, based on all
$\epm$ data including more recent BaBar and KLOE data, yields
$a_\mu^{\mathrm{had,LO}}[e]=690.75(4.72)\power{-10}$ ($\epm$ based),
while including $\tau$ data we find
$a_\mu^{\mathrm{had,LO}}[e,\tau]=690.96(4.65)\power{-10}$
($\epm$+$\tau$ based). This backs the $\sim$3$\,\sigma$ deviation
between $\amu^{\mathrm{experiment}}$ and $\amu^{\mathrm{theory}}$. For
the $\tau$ di-pion branching fraction we find
$B^{\mathrm{CVC}}_{\pi\pi^0}=25.20\pm0.0.17\pm0.28$ from $\epm$+CVC,
while $B_{\pi\pi^0}=25.34\pm0.0.06\pm0.08$ is evaluated directly from
the $\tau$ spectra.
\end{abstract}

\end{frontmatter}

\renewcommand{\thefootnote}{\arabic{footnote}}
\setcounter{footnote}{0}

\section{Introduction}
Isovector data for the pion form factor obtained from hadronic
$\tau$-decay spectra can be compared with the mixed
isovector-isoscalar data measured in the $\epm$ channel by means of
theory input~\cite{Tsai}. In particular, we need some model in order
to be able to disentangle $\rho-\omega$ mixing as well as other
isospin breaking (IB) effects. The general problem in confronting measured
quantities like $|F^{(\tau)}_\pi(s)[I=1]|^2$ and
$|F^{(e)}_\pi(s)|^2=|F^{(e)}_\pi(s)[I=1]+F^{(e)}_\pi(s)[I=0]|^2$ is
the fact that the latter object is subject to quantum interference
between the two amplitudes and in general may not be well
approximated by
$|F^{(e)}_\pi(s)[I=1]|^2+|F^{(e)}_\pi(s)[I=0]|^2$. Without a
specific model for the complex amplitudes one cannot get the precise
relationship.

Commonly, pion form factors measured in the neutral
channel in $\eepp$ and in the charged channel in $\tau^- \to \nu_\tau
\pi^- \pi^0$ decay (or its charge conjugate) are parametrized by an
extended Gounaris-Sakurai (GS) formula
\be
F_\pi(s)=\frac{\mathrm{BW}^\mathrm{GS}_{\rho(770)}(s)\cdot
\left(1+\delta \frac{s}{M_\omega^2}\mathrm{BW}_{\omega}(s)\right)
+\beta\: \mathrm{BW}^\mathrm{GS}_{\rho(1450)}(s)
+\gamma\: \mathrm{BW}^\mathrm{GS}_{\rho(1700)}(s)}{1+\beta+\gamma}\cs
\label{GSform}
\ee
which results as a sum of mixing isovector states, each described
by a Breit-Wigner (BW) type of amplitude.  The pion form factor is related
to the corresponding cross section by\footnote{QED corrections to $\eepp$
have been summarized in~\cite{Axeletal02,Actis:2010gg}. They will not be considered
in the following and we assume them to be taken into account in the
extraction of the form factor from the experiments.}
\ba
\sigma(\eepp)=\frac{\pi
\alpha^2}{3}\,\frac{\beta_\pi^3}{s}\,|F^{(e)}_\pi(s)|^2=\frac{4\pi\alpha^2}{s}\,v_0(s),
\label{Xsection}
\ea
for point-like pions $F_\pi(s) \equiv 1$, where $\beta_\pi$ is the
pion velocity in the c.m. frame: $\beta_\pi=\sqrt{1-4m_\pi^2/s}$. The
spectral function $v_i(s)$ is related to the form factor by
\ba
v_i(s)=\frac{\beta_i(s)}{12 \pi}\,|F^{(i)}_\pi(s)|^2\semis (i=0,-)
\leftrightarrow (e,\tau)\cs
\label{vspectral}
\ea
for the neutral (0) $\epm$- and charged (-) $\tau$-channel. The
spectral function $v_-(s)$ can be measured very precisely in $\tau$-decay:
\ba
\frac{1}{\Gamma}\frac{\D \Gamma}{\D s}(\tau^- \to \nu_\tau \pi^- \pi^0)
=\frac{6|V_{ud}|^2\,S_{\rm EW}}{m_\tau^2}\frac{B_e}{B_{\pi\pi}}\,\left(1-\frac{s}{m_\tau^2}\right) 
\left(1+\frac{2s}{m_\tau^2}\right)\,v_-(s)\cs
\ea
with $m_\tau=(1776.84\pm0.17)~\mv$ the $\tau$ mass,
$|V_{ud}|=0.9418\pm0.00019$ the CKM matrix element,
$B_e=(17.818\pm0.032)\,\%$ the electron branching fraction,
$B_{\pi\pi}=(25.51\pm 0.09)\,\% $ the di-pion branching fraction and
$S_{\mathrm{EW}}=1.0235\pm 0.0003$ the short distance electroweak
correction.

Note that a single standard Breit-Wigner resonance yields
\bea
|F_\pi(s)|^2=\frac{36}{\alpha^2}\,
\frac{\Gamma(\rho \to \epm)}{\beta_\pi^3 \Gamma(\rho \to \pipi)}\,
\frac{s}{\mr}\,\frac{s\Gamma_\rho^2}{(s-\mr)^2+\mr \Gamma_\rho^2}\epo
\eea
Denoting the $\gamma - \rho$ transition coupling by $e\mr/g_\rho$ the
branching fraction at resonance reads
\bea
R_\rho\doteq\frac{\Gamma(\rho \to \epm)}{\Gamma(\rho \to \pipi)} =
\frac{\alpha^2}{36}\,\left(\frac{g_\rho}{g_{\rho\pi\pi}}\right)^2\,
\left(\frac{M_\rho}{\Gamma_\rho}\right)^2\, \beta_\rho^3\cs
\eea
with $\beta_\rho\doteq\beta_\pi(s=\mr)$. In the case of complete
$\rho$ dominance $g_\rho=g_{\rho\pi\pi}$\footnote{At resonance
the single BW pion form factor is given by
\bea
|F_\pi(\mr)|^2=\frac{36}{\alpha^2}\,
\frac{\Gamma_{ee}}{\beta_\rho^3 \Gamma_{\pi\pi}}\cs
\eea
and for PDG values of the parameters yields $|F_\pi(\mr)|^2\approx 39$
a reasonable value (see below).}.

The GS formula (\ref{GSform}) also describes the charged isovector
channel provided $\delta=0$, since there is no charged
version of the $\omega$. In the neutral channel the GS formula does
not fully include \rogami, which is known since the
early 1960's, when the $\rho$ had been discovered. A direct consequence
of \rogami is the vector meson dominance (VMD)
model characterized by an effective Lagrangian~\cite{VMD-1}
\ba
{\cL}_{\gamma \rho}=-\frac{e\,M_\rho^2}{g_\rho}\,\rho_\mu A^\mu\epo
\ea
However, this form does not preserve electromagnetic gauge invariance
and the photon would acquire a mass unless we add a photon mass counterterm
to the Lagrangian which is fine tuned appropriately.
The pion form factor here takes the form
\ba
F_\pi(s)=-\frac{M_\rho^2}{s-M_\rho^2}\, \frac{g_{\rho\pi\pi}}{g_\rho}
\label{VMDnaive}
\ea
and the condition of electromagnetic current conservation $F_\pi(0)=1$
is satisfied only if $g_{\rho\pi\pi}=g_\rho$, which is called
universality condition and corresponds to complete $\rho$
dominance. In fact electromagnetic gauge invariance can be implemented
by writing the effective VMD Lagrangian in the form~\cite{VMD-2}
\ba
{\cL}_{\gamma \rho}=\frac{e}{2\,g_\rho}\,\rho_{\mu\nu} F^{\mu\nu}\comas
\label{Lrhogamma}
\ea
in terms of the field strength tensors. As it satisfies gauge
invariance, the form factor calculated here reads
\ba
F_\pi(s)=1-\frac{s}{s-M_\rho^2}\, \frac{g_{\rho\pi\pi}}{g_\rho}
\ea
and satisfies the current conservation condition $F_\pi(0)=1$ in any
case, irrespective of the universality constraint
$g_{\rho\pi\pi}=g_\rho$ (for a recent discussion also
see~\cite{Zerwekh06}).  Obviously, this simple model is not able to
describe the pion form factor measured in $\eepp$ at low energies,
unless we take into account energy dependent finite widths effects of
the $\rho$ as it is done in the GS model~\cite{GS68}\footnote{Other
models have been reviewed and investigated recently with emphasis on
$\rho-\omega$ mixing in Ref.~\cite{Wolfe:2009ts}. Frequently used
descriptions of the low energy $\pi\pi$ form factor include the
ChPT-based Guerrero-Pich formulation~\cite{Guerrero:1997ku}, the
Leutwyler-Colangelo approach~\cite{LeCo02}, the resonance Lagrangian
approach~\cite{EckerCPT} (see e.g.~\cite{Ivashyn:2006gf}) or the
related Hidden Local Symmetry (HLS) model as applied
in~\cite{Benayoun07,Benayoun10}, and the phenomenological
K\"uhn-Santamaria (KS) model~\cite{Kuhn:1990ad}. As we will see our
model is closely related to the GS model, and we adopt the latter for
comparisons and fits.}. The energy dependence of the $\rho$-width has
to reflect the off-shell $\rho^*
\to \pi\pi$ process. So we have to model effectively a
``rho-pion-photon'' system, discarding the $\omega$ and its mixing
with the $\rho$, which is well understood and will be taken into
consideration in a second step.  Our focus here is to work out the
difference in the relation between the charged channel and the neutral
channel, which results from the \rogami. The latter like
$\rho^0-\omega$ mixing, has no counterpart in the charged channel.
The purpose of this study is to understand better the discrepancy
between $\tau$ and $\epm$ di-pion spectra, which has been clearly established
in~\cite{DEHZ03} under the assumption that all possible IB corrections
were accounted for. More recent data form Belle~\cite{Belle08}
and KLOE~\cite{KLOE08,KLOE10}, and applying improved IB corrections, 
confirmed a significant discrepancy~\cite{Davier:2009ag}. Although the new
$\pi^+\pi^-$ spectrum from BaBar~\cite{BABAR09}, measured via the
radiative return mechanism, is closer to the corresponding spectra
obtained from $\tau$-decays, a discrepancy persists.

\section{A $\rho-\gamma$ mixing model and related self-energy effects}

As already said, the VMD ansatz has to be replaced by a more realistic
model which must  take into account
\bit
\item the finite $\rho$-width, related to its decay $\rho \to \pi^+\pi^-$,
\item the $\rho-\gamma$ mixing, which leads to non-diagonal
      propagation of the $\rho-\gamma$ system, and
\item the $\rho-\omega$ mixing, which we will consider in a second
      step.
\eit
This has to be implemented in an appropriate effective field theory (EFT).
In a first step we consider the interaction of the $\rho$ with the
pions together with their electromagnetic interaction, assuming the
pions to by point-like (scalar QED). As suggested long ago by
Sakurai~\cite{Sakurai60}, the $\rho$ may be treated as a massive gauge
boson. The effective Lagrangian thus reads
\ba
\cL=\cL_{\gamma \rho}+\cL_\pi \semis
\cL_\pi=D_\mu \pi^+ D^{+\mu} \pi^- -m_\pi^2 \pi^+\pi^- \semis
D_\mu = \partial_\mu -\I \,e\,A_\mu -\I\,g_{\rho\pi\pi}\rho_\mu \epo
\label{ModelSimple}
\ea
The corresponding Feynman rules in momentum space are
\begin{center}
\begin{tabular}{ccccccc}
$A^\mu \pi\pi        $&$\corr $&$ -\I\,e\,(p+p')^\mu $&$\semis$&
$\rho^\mu \pi\pi     $&$\corr $&$ -\I\,g_{\rho\pi\pi}\,(p+p')^\mu $\\
$A^\mu A^\nu\pi\pi   $&$\corr $&$ 2\,\I\,e^2\, g^{\mu\nu} $&$\semis$&
$\rho^\mu\rho^\nu\pi\pi    $&$\corr $&$ 2\,\I\,g_{\rho\pi\pi}^2\, g^{\mu\nu} $\\
$A^\mu\rho^\nu\pi\pi $&$\corr $&$ 2\,\I\,e\,g_{\rho\pi\pi}\, g^{\mu\nu} $&$\semis$&
$A^\mu \rho^\nu $&$\corr$&$ -\I\,e/g_\rho\,(p^2\,g^{\mu\nu}-p^\mu p^\nu)\epo$
\end{tabular}
\end{center}
The model should be understood as a simplified version of the better
justified effective resonance Lagrangian approach~\cite{EckerCPT},
which extends the chiral structure of low energy QCD (chiral
perturbation theory) to include spin 1 resonances in a consistent
way. A variant is the HLS model, which in the same context has been
applied to investigate the $(\rho,\omega,\phi)$ mixing effects
in~\cite{Benayoun07}. Actually in~\cite{Benayoun07} too, $V-\gamma$
($V=\rho,\omega,\phi$) mixing amplitudes have been included
(more on that below). The main difference to the GS model is that we
take our EFT Lagrangian serious in the sense that we include all
relevant contributions to $\eepp$, while in the GS model some of the
contributions have been neglected. In fact the GS model is incomplete
in the sense of a quantum field theory.
%\clearpage

In sQED the contribution of a pion loop to the photon vacuum
polarization is given diagrammatically by\\
\begin{figure}[h]
\centering
\includegraphics[height=1.25cm]{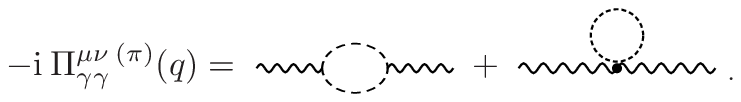}
\end{figure}

\noi
and one then obtains the  bare $\gamma\,-\,\rho$ transverse self-energy functions
\ba
\Pi_{\gamma \gamma}&=&\frac{e^2}{48 \pi^2}\,f(q^2)\comas
\Pi_{\gamma \rho}=\frac{e g_{\rho\pi\pi}}{48 \pi^2}\,f(q^2)\;\: \mathrm{ \ and \ }\;\;
\Pi_{\rho \rho}=\frac{g^2_{\rho\pi\pi}}{48 \pi^2}\,f(q^2)\comas
\label{SEsgr}
\ea
where
\ba
f(q^2)&\equiv&q^2\,h(q^2)=
\left(B_0(m_\pi,m_\pi;q^2)\,(q^2-4\,m_\pi^2)-4\,A_0(m_\pi)-4\,m_\pi^2+
\frac{2}{3}q^2\right)\cs
\ea
in terms of the standard scalar one-loop integrals $A_0(m)$ and $B_0(m,m;s)$~\cite{FJbook}.
Explicitly\footnote{The standard Gounaris-Sakurai
parametrization differs from our sQED model and utilizes
$h(q^2)=-2\,(1-y)^2\,G(y)$ which for $q^2 \to 0$ behaves as $h(q^2)\to
2y=8m_\pi^2/q^2$ i.e. $q^2h(q^2) \to 8m_\pi^2$, which in $D_{\gamma
\gamma}$ represents a non-vanishing photon mass. While the constant
terms in (\ref{hfun}) drop out by renormalization, the $+2(1-y)$ term
is required by electromagnetic gauge invariance and in fact renders
$h(q^2) \to \mathrm{const.}$ regular in the static limit.}, in the
\MSb scheme ($\mu$ the \MSb renormalization scale)
\ba
h(q^2)\equiv f(q^2)/q^2
&=&2/3+2\,(1-y)-2\,(1-y)^2\,G(y)+\ln \frac{\mu^2}{m_\pi^2}\cs
\label{hfun}
\ea
where $y=4m_\pi^2/s$ and $G(y)=\frac{1}{2\beta_\pi}\,(\ln
\frac{1+\beta_\pi}{1-\beta_\pi}-\I\,\pi)$, for $q^2 > 4 \,m_\pi^2$.  Note that
all components of the ($\gamma$, $\rho$) 2$\times$2 matrix propagator
are proportional to the same function $f(q^2)$.  The renormalization
conditions are such that the matrix is diagonal and of residue unity
at the photon pole $q^2=0$ and at the $\rho$ resonance $s=M_\rho^2$,
hence the renormalized self-energies read (see e.g.~\cite{TASI})
\ba
\Pi^\mathrm{ren}_{\gamma \gamma}(q^2)&=&\Pi_{\gamma
\gamma}(q^2)-q^2\,\Pi'_{\gamma \gamma}(0) \doteq q^2\,\Pi^\mathrm{'ren}_{\gamma \gamma}(q^2)\\
\label{grren}
\Pi^\mathrm{ren}_{\gamma \rho}(q^2)&=&  \Pi_{\gamma \rho}(q^2)
-\frac{q^2}{M_\rho^2}\,\Repa\Pi_{\gamma \rho}(M_\rho^2)\\
\label{rrren}
\Pi^\mathrm{ren}_{\rho \rho}(q^2)&=&	   \Pi_{\rho \rho}(q^2)-\Repa
\Pi_{\rho \rho}(M_\rho^2)-(q^2-M_\rho^2)\,\Repa \frac{\D\Pi_{\rho
\rho}}{\D s}(M_\rho^2)
\ea
where $\Pi_{\gamma \gamma}(0)=\Pi_{\gamma \rho}(0)=\Pi_{\rho \rho}(0)=0$ and 
$\Pi'_{\gamma \gamma}(q^2)=\Pi_{\gamma \gamma}(q^2)/q^2$,
has been used. Note, that the tree level mixing term in the Lagrangian
contributes to the bare $\gamma \rho$ self-energy as
$\Pi^{(0)}_{\gamma \rho} =q^2 (e/g_\rho)$, which does not affect the renormalized
self energies, however. In particular, $\delta \Pi^\mathrm{ren}_{\gamma \rho}=
q^2\,\frac{e}{g_\rho}-\frac{q^2}{\mr}\,\mr \frac{e}{g_\rho}=0$. The
$\rho$ wave function renormalization reads
$Z_\rho=1/(1+\frac{\D \Pi_{\rho \rho}}{\D s}(s=\mr))$ with 
\be
\frac{\D \Pi_{\rho \rho}}{\D s}(s=\mr)=\frac{g_{\rho\pi\pi}^2}{48\,\pi^2}\,
\left\{8/3-\beta_\rho^2
\,\left[1+ (3-\beta_\rho^2)\,\frac{1}{2\,\beta_\rho}
\ln \left(\frac{1+\beta_\rho}{1-\beta_\rho}\right)\right]\right\}\epo
\ee
Numerically,  $Z_\rho \simeq  1.1289$ at $\mu=m_\pi$.

It is crucial to observe that vacuum polarization
effects affect mass renormalization of the $\rho$ as well as $\gamma$
- $\rho$ mixing, in spite of the fact that photon vacuum polarization
has to be subtracted in the definition of $F_\pi$. In other words, in sQED we would
still have $F_\pi(s)=1$ in (\ref{Xsection}) while vacuum polarization
is absorbed into a running fine structure constant
\ba
\alpha \to \alpha(s)=\frac{\alpha}{1+\Pi^\mathrm{'ren}_{\gamma \gamma}(s)}\cs
\ea
which mean that in calculating $F_\pi(s)$ we have to multiply the result
by $1+\Pi^\mathrm{'ren}_{\gamma \gamma}(s)$.

A convenient representation of $\Pi^\mathrm{ren}_{\rho \rho}$ is given
by
\ba
\Pi^\mathrm{ren}_{\rho \rho}(s) &=& \frac{\Gamma_\rho}{\pi M_\rho\,\beta_\rho^3}\,
\left\{s\,\left(h(s)-\Repa h(M_\rho^2)\right)-
(s-M_\rho^2)\,M_\rho^2\,\left.\Repa \frac{\D h}{\D s}\right|_{s=M_\rho^2}\right\}
\label{Pirhorep}
\ea
with
\ba
s\,\frac{\D h}{\D s}(s) = 3\,y-1-3\,y\,(1-y)\,G(y)\epo
\ea
In particular\footnote{In contrast to sQED, in the standard GS formula
$$h(s)=-2(1-y)^2\,G(y)\semis s\,\frac{\D h}{\D s}(s) = y-1-3\,y\,(1-y)\,G(y)\cs$$
which is singular for $s \to 0$. The first term in (\ref{Pirhorep})
in this case yields a finite contribution $sh(s)\to 8m_\pi^2$ and thus
\bea
\Pi^\mathrm{ren}_{\rho \rho}(0) = \frac{M_\rho \Gamma_\rho}{\pi\,\beta_\rho^3}\,
\left(\frac{8m_\pi^2}{M_\rho^2}+M_\rho^2\,\left.\frac{\D h}{\D s}\right|_{s=M_\rho^2}\right)=
\frac{M_\rho \Gamma_\rho}{\pi\,\beta_\rho^3}\,
\left(3y_\rho-1-3\,y_\rho\,\beta_\rho^2\,G(y_\rho)\right)\cs
\eea
i.e., $\Pi^\mathrm{ren}_{\rho \rho}(0)\equiv -d\,\Gamma_\rho\,M_\rho$ is actually not modified,
in spite of lacking manifest gauge invariance.}:
\ba
\Pi^\mathrm{ren}_{\rho \rho}(0) &=& \frac{M_\rho \Gamma_\rho}{\pi\,\beta_\rho^3}\,
M_\rho^2\,\left.\frac{\D h}{\D s}\right|_{s=M_\rho^2}=
\frac{M_\rho \Gamma_\rho}{\pi\,\beta_\rho^3}\,
\left(3y_\rho-1-3\,y_\rho\,\beta_\rho^2\,G(y_\rho)\right)\semis y_\rho=4m_\pi^2/M_\rho^2\epo
\ea
Without mixing, pion production mediated by the $\rho$ resonance,
yields the GS type pion form factor, normalized to $F_\pi(0)=1$,
\ba
F^{\mathrm{GS}}_\pi(s)=\frac{-M_\rho^2+\Pi^\mathrm{ren}_{\rho \rho}(0)}{s-M_\rho^2+\Pi^\mathrm{ren}_{\rho \rho}(s)}\epo
\label{GSfin}
\ea
The renormalized mixing self-energy may be written in a form
\ba
\Pi^\mathrm{ren}_{\gamma \rho}(s) &=& \frac{eg_{\rho \pi\pi}}{48 \pi^2}\,
\left\{s\,\left(h(s)-\Repa h(M_\rho^2)\right)\right\}\epo
\label{Pigrrep}
\ea

Note that while the inverse propagator matrix is diagonal at the two
propagator poles, off the poles it is not diagonal. This is the main
effect we are going to discuss now\footnote{These effects are very
similar to $Z^0-\gamma$ mixing~\cite{Glashow61} which has been
investigated theoretically as well as experimentally at LEP with high
precision (see Refs.~\cite{LEP01,FJ91} and references
therein). Typically, these effects at low energy or near the $Z$ pole
are expected to be small because of the smallness of the
electromagnetic fine structure constant, and at the $Z$ resonance,
because of the large mass and very small width of the $Z^0$ boson. In
case of the $\rho$ witch's mass lies not very far above the hadronic
$\pi\pi$-threshold (which is very low by the fact that pions are quasi
Nambu-Goldstone bosons) and due to the relatively large (hadronic)
width we expect corresponding mixing effects to be much more
relevant. In the charged channel, in principle, there is $W^\pm -
\rho^\pm$ mixing, with some effective $W^+\rho^- + \mathrm{h.c.}$
coupling term. However, this produces a negligible effect because the
$W$ propagator pole is far away from the $\rho$ propagator pole and
from the energy range of interest. In fact the mixing matrix,
diagonalized at the $\rho$-pole,  
remains essentially diagonal in the whole range of
interest ($<$ 2 GeV).}.

The propagators are obtained by inverting the symmetric $2\times2$ self energy
matrix
\ba
\hat{D}^{-1}= \left ( \begin{array}{cc}
                      q^2+\Pi_{\gamma \gamma}(q^2) & \Pi_{\gamma \rho}(q^2) \\
                      \Pi_{\gamma \rho}(q^2) & q^2-\mr + \Pi_{\rho\rho} (q^2)
                                    \end{array} \right )
\label{invprop22}
\ea
with the result:
\ba
D_{\gamma \gamma} &=& \frac{1}{q^2+\Pi_{\gamma \gamma}(q^2)
-\frac{\Pi_{\gamma \rho}^2 (q^2)}{q^2-\mr +\Pi_{\rho\rho}(q^2)}}
\simeq \frac{1}{q^2+\Pi_{\gamma \gamma}(q^2)}  \crn
D_{\gamma \rho} &=& \frac{-\Pi_{\gamma \rho}(q^2)}{(q^2+\Pi_{\gamma \gamma}(q^2))
(q^2-\mr +\Pi_{\rho\rho}(q^2))-\Pi_{\gamma \rho}^2 (q^2)}
\simeq \frac{-\Pi_{\gamma \rho} (q^2)}{(q^2+\Pi_{\gamma \gamma}(q^2))
(q^2-\mr +\Pi_{\rho\rho}(q^2))}  \crn
D_{\rho\rho} &=& \frac{1}{q^2-\mr +\Pi_{\rho\rho}(q^2)
-\frac{\Pi_{\gamma \rho}^2 (q^2)}{q^2+\Pi_{\gamma \gamma}(q^2)}}
\simeq \frac{1}{q^2-\mr +\Pi_{\rho\rho}(q^2)}\;.
\label{invpropterms}
\ea
These expressions sum correctly all the irreducible self-energy
bubbles\footnote{It is of curse well known that this Dyson summation
is crucial for a proper description of the particle/resonance
structure in particular near the poles, where naive perturbation
theory in any case breaks down.}. The approximations indicated are the
one-loop results. The extra terms are higher order contributions and
are particularly relevant near the resonance, characterized by the
location $s_P$ of the pole of the propagator, which is given by the
zero of the inverse propagator:
\begin{equation}
\label{Zgamma}
s_P - m_{\rho^0}^2 - \Pi_{\rho^0\rho^0}(s_P)
- \frac{\Pi^2_{\gamma \rho^0}(s_P)}{s_P-\Pi_{\gamma \gamma}(s_P)} = 0\cs
\end{equation}
with $s_P=\tilde{M}^2_{\rho^0}$ complex.  The usual (no mixing)
considerations in determining the physical mass and width of a
resonance remain true if we denote the self-energies by $\Pi_V$
($V=\rho^0,\rho^\pm$) with
$$\Pi_{\rho^\pm}(p^2,\cdots)=\Pi_{\rho^+\rho^-}(p^2,\cdots)$$ and
$$\Pi_{\rho^0}(p^2,\cdots)= \Pi_{\rho^0\rho^0}(p^2,\cdots)
+ \frac{\Pi^2_{\gamma \rho^0}(p^2,\cdots)}{p^2-\Pi_{\gamma
\gamma}(p^2,\cdots)}\;.$$ Thus
the location of the pole may be written as
\begin{equation}
\tilde{M}^2 - m^2 + \Pi(\tilde{M}^2,m^2,\cdots) = 0,
\label{pole}
\end{equation}
for both the $\rho^\pm$ and the $\rho^0$, where
\bea
\tilde{M}^2_{\rho} \equiv \left(q^2\right)_{\rm pole} =\mr-\I\;M_{\rho}\; \Gamma_{\rho}
\eea
is characterized by mass and width of the $\rho$\footnote{For $\gamma$
- $Z^0$ mixing in the electroweak Standard Model explicit
results up to two loops have been worked out in~\cite{GBM2}.}. Note that the
imaginary part of the self-energy function is energy dependent, which
implies an energy dependent width, of course with the correct
phase-space behavior of $\rho \to \pi\pi$ decay.

\begin{figure}[t]
\centering
\includegraphics[height=6cm]{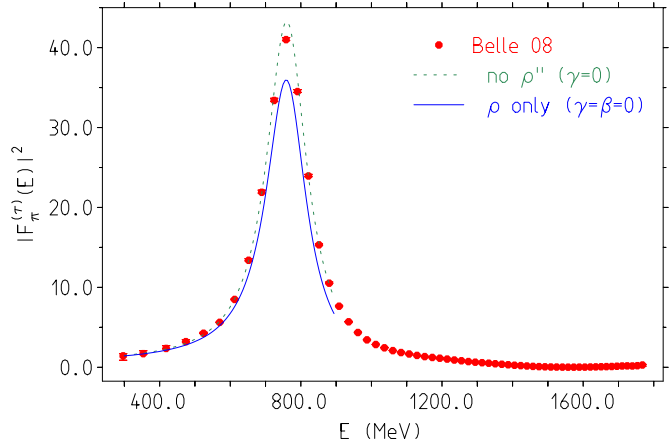}
\caption{GS fits of the Belle data and the effects of including
higher states $\rho'$ and $\rho''$ at fixed $M_\rho$ and
$\Gamma_\rho$. Doubts are in order whether the higher resonances
really affect the $\rho$ resonance in the way suggested by the
commonly adopted GS parametrization.}
\label{fig:GS1}
\end{figure}

How do off-diagonal elements of the $\gamma\,-\,\rho$ propagator
affect the line-shape of the $\rho$? We assume we know the $\rho$ mass
$M_\rho$ and the $\rho$ width $\Gamma_\rho$ for the unmixed $\rho$ as it is seen
e.g. in the isovector $\tau$ decay spectra, i.e. we compare the result
with a charged $\rho^\pm$ assuming equal mass and width. We therefore
compare result first with the Belle data \cite{Belle08}. Of course our
model does not fit the data, because a more sophisticated extended
Gounaris-Sakurai model (\ref{GSform}) has been used to extract
the $\rho$ parameters. If we switch off the contributions from
$\rho'$ and $\rho''$ by setting $\gamma=0$ and $\beta=\gamma=0$
we observe a substantial change in $F_\pi(s)$ as illustrated in
Fig.~\ref{fig:GS1}.  Note that in an EFT one would expect the heavier
states to decouple, while in the GS type modeling the low energy tail
is normalized away by the $1+\beta+\gamma$ normalization factor. In
field theory in place of this normalization a factor
$s/M_{\mathrm{res}}^2$ would imply automatic decoupling. But
that is not the way mass and width of the $\rho$ are determined
usually.  Evidently, in the GS model, in the $\rho$-region the higher
resonances serve as a continuum background without which good fits in
general are not possible. So if we stick with our simplified model we
cannot expect to get a good representation of the data without
corresponding extensions. On the other hand, the simplified model
allows us to work out more clearly the effect of $\gamma$ - $\rho$
mixing.
Above we have diagonalized the mixing propagator matrix at the poles,
this allows to make precise the meaning of mass and width of the heavy
unstable state. This is achieved if renormalized mixing self-energy is
given by
\ba
\Pi_{\gamma \rho\;ren}(q^2)=\Pi_{\gamma \rho}(q^2)-\Pi_{\gamma \rho}(0)-
\frac{q^2}{\mr}\left(Re \Pi_{\gamma \rho}(\mr)-\Pi_{\gamma \rho}(0)\right)\;.
\label{renmix}
\ea
This can be achieved by two subsequent transformations of the bare fields:
\begin{itemize}
\item[i)] Infinitesimal (perturbative) rotation
\bea
\left( \begin{array}{c} A_b \\
                        \rho_b \end{array} \right)\;=\;
\left( \begin{array}{rr} 1 & -\Delta_0 \\
                        \Delta_0 & 1  \end{array} \right)
\left( \begin{array}{c} A' \\
                        \rho' \end{array} \right)
\label{fr1}
\eea
          diagonalizing the mass
          matrix at one-loop (n+1-loop) order given that the mass matrix
          has been diagonalized at tree (n-loop) level.
\item[ii)] Upper diagonal matrix wave function renormalization
           inducing a kinetic mixing term (this cannot be done by an
           orthogonal transformation)
\bea
\left( \begin{array}{c} A' \\
                        \rho' \end{array} \right)\;=\;
\left( \begin{array}{rr} \sqrt{Z_{\gamma}} & -\Delta_\rho  \\
                            0      & \sqrt{Z_\rho} \end{array} \right)
\left( \begin{array}{c} A_r \\
                        \rho_r \end{array} \right)
\label{fr2}
\eea
          which allows to normalize the residues to one for the $\gamma$- and
          $\rho$-propagator, respectively, and to shift to zero the mixing propagator at the $\rho$-pole.
\end{itemize}
Thus the relationship between the bare and the renormalized (LSZ) fields is
(expanded to linear order)
\ba
A_b&=&\sqrt{Z_{\gamma}}A_r -\:(\Delta_\rho+\Delta_0)\;\rho_r \crn
\rho_b&=&\sqrt{Z_{\rho}}\rho_r +\Delta_0\; A_r\;,
\ea
generalizing the usual multiplicative field renormalization
represented by the first term for both fields. The counter-terms
$\Delta_0$ and $\Delta_\rho$ are determined by the condition
(\ref{renmix})
\ba
\Delta_0 &=&\frac{\Pi_{\gamma \rho}(0)}{\mr} \crn
\Delta_\rho &=&\frac{Re \Pi_{\gamma \rho}(\mr)-\Pi_{\gamma \rho}(0)}{\mr}\epo
\ea
For our model $\Delta_0=0$ and $\Delta_\rho=e/g_\rho$ to leading
order.  The field transformations of course induce mixing counter
terms at the vertices, which are absorbed into the definition of the
physical couplings. In principle, this non-symmetric transformation
only affects the bookkeeping such that the propagator pole structure
becomes obvious. It does not change the value of the functional
integral i.e. the mixing counter terms cancel in the interior of
Feynman diagrams, unless the photon and/or the rho are involved as
external fields (states).

As a consequence of the diagonalization the physical $\rho$ acquires a direct
coupling to the electron: starting as usual from the bare Lagrangian
\ba
\cL_{\mathrm{QED}}=\bar{\psi}_e \gamma^\mu (\partial_\mu-\I\,e_b\,A_{b\mu})\,\psi_e
\ea
we obtain
\ba
\cL_{\mathrm{QED}}=\bar{\psi}_e \gamma^\mu
(\partial_\mu-\I\,e\,A_{\mu}+\I\,g_{\rho ee}\rho_\mu)\,\psi_e
\label{Leegr}
\ea
with $g_{\rho ee}=e\,(\Delta_\rho+\Delta_0)$, where in our case $\Delta_0=0$.

\begin{figure}[h]
\centering
\includegraphics[height=1.5cm]{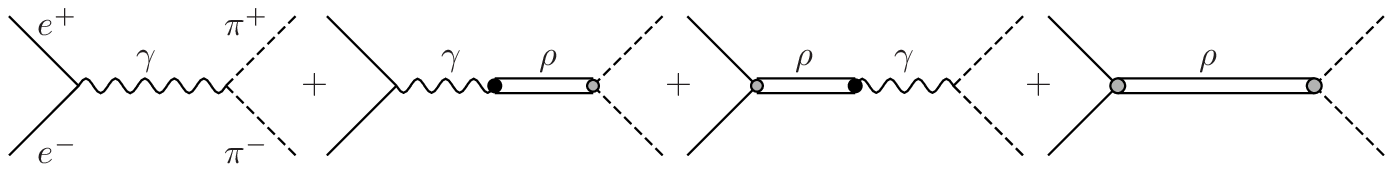}
\caption{Diagrams contributing to the process $\eepp$.}
\label{fig:eepipi} 
\end{figure}

The $\eepp$ matrix element in sQED is given by
\be
\cM =-\I\,e^2\,\bar{v} \gamma^\mu u \,(p_1-p_2)_\mu\,F_\pi(q^2)
\ee
with $F_\pi(q^2)=1$. 
In our extended VMD model we have the four terms
shown in Fig.~\ref{fig:eepipi} and thus 
$$F_\pi(s) \propto  e^2\,D_{\gamma\gamma}+ eg_{\rho\pi\pi}\,D_{\gamma \rho}-
g_{\rho ee}e D_{\rho \gamma}
-g_{\rho ee} g_{\rho\pi\pi} \,D_{\rho\rho}\,,$$
where the first term properly normalized must be unity\footnote{Note
that the conserved vector current (CVC) condition $F_\pi(0)=1$ in our
model is given and saturated by the sQED term $D_{\gamma
\gamma}$ alone,
while in the GS model $F_\pi(0)=1$ is imposed by force on the
term $D_{\rho\rho}$, the only one present in the GS case.}. Thus
\be
F_\pi(s) =  \left[e^2\,D_{\gamma\gamma}+ e\,(g_{\rho\pi\pi}-g_{\rho ee})\,D_{\gamma \rho}-
g_{\rho ee} g_{\rho\pi\pi} \,D_{\rho\rho}\right]/\left[e^2\,D_{\gamma\gamma}\right]\epo
\label{fpidecomp}
\ee
Note the sign of the induced coupling $g_{\rho e e}$ in (\ref{Leegr}),
which leads to the signs as given in (\ref{fpidecomp}).  Typical
couplings read $g_{\rho\pi\pi\,\mathrm{bare}} = 5.8935$,
$g_{\rho\pi\pi\, \mathrm{ren}} = 6.1559$, $g_{\rho ee} =  0.018149$ and
$x=g_{\rho\pi\pi}/g_\rho=   1.15128$.

\begin{figure}
\centering
\includegraphics[height=5cm]{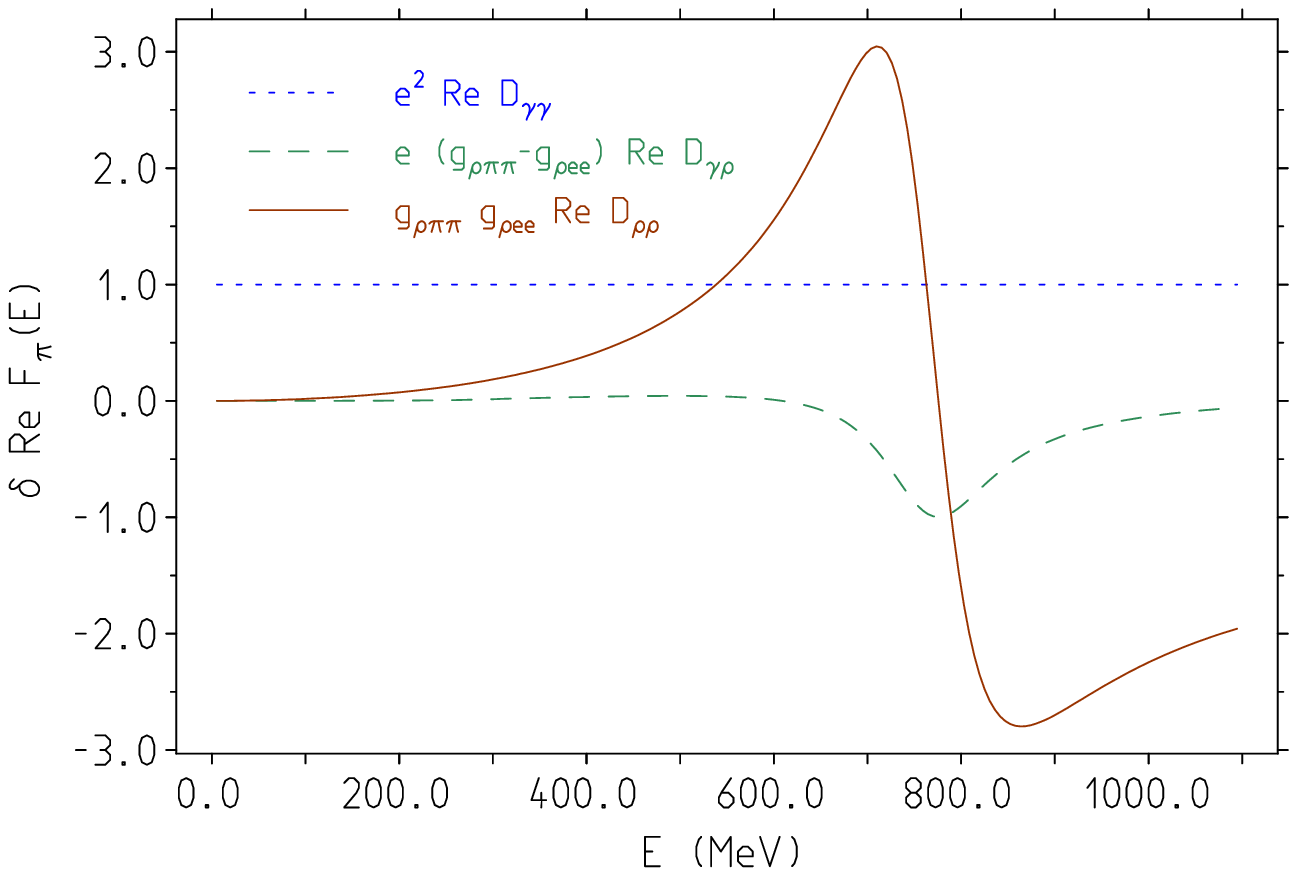}
\includegraphics[height=5cm]{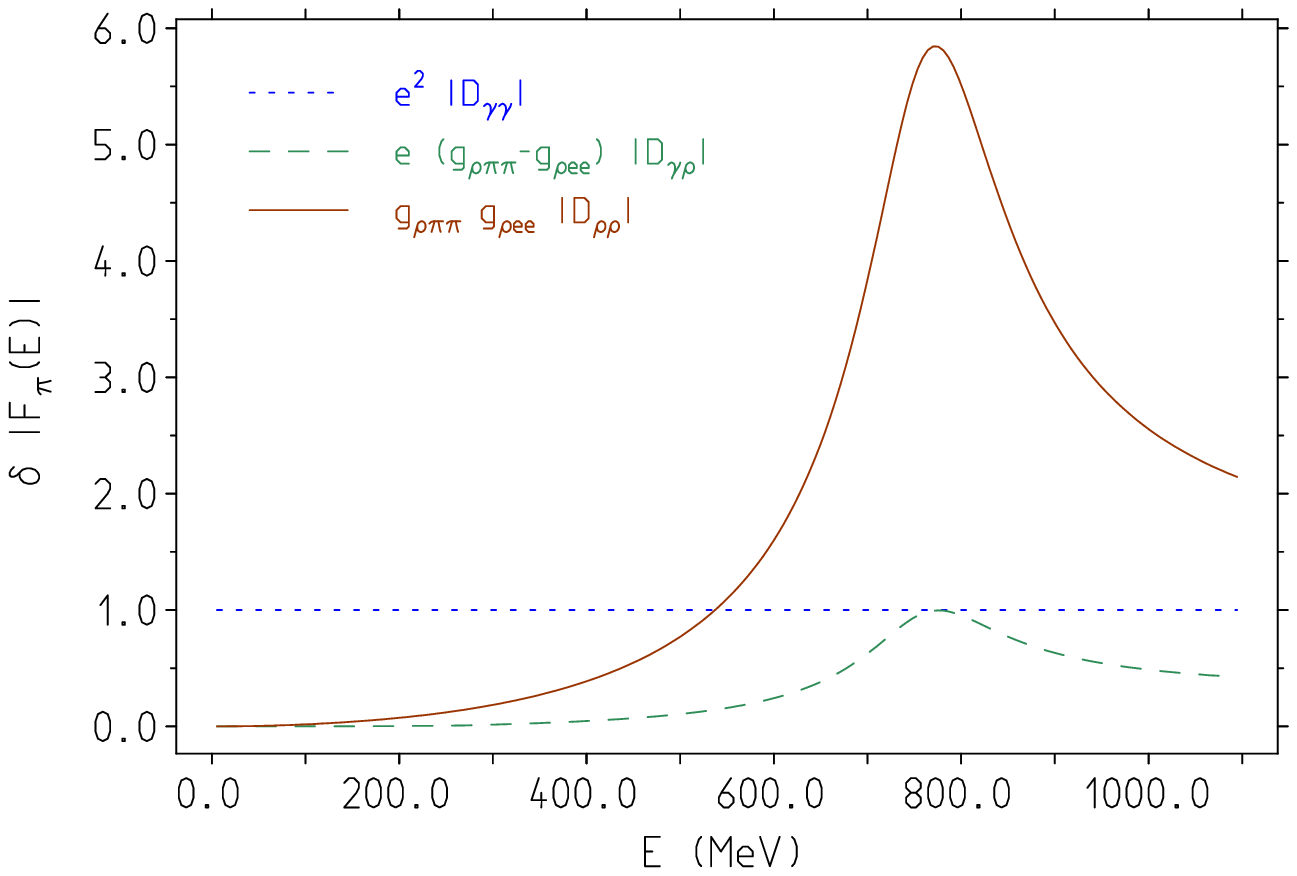}\\
\includegraphics[height=5cm]{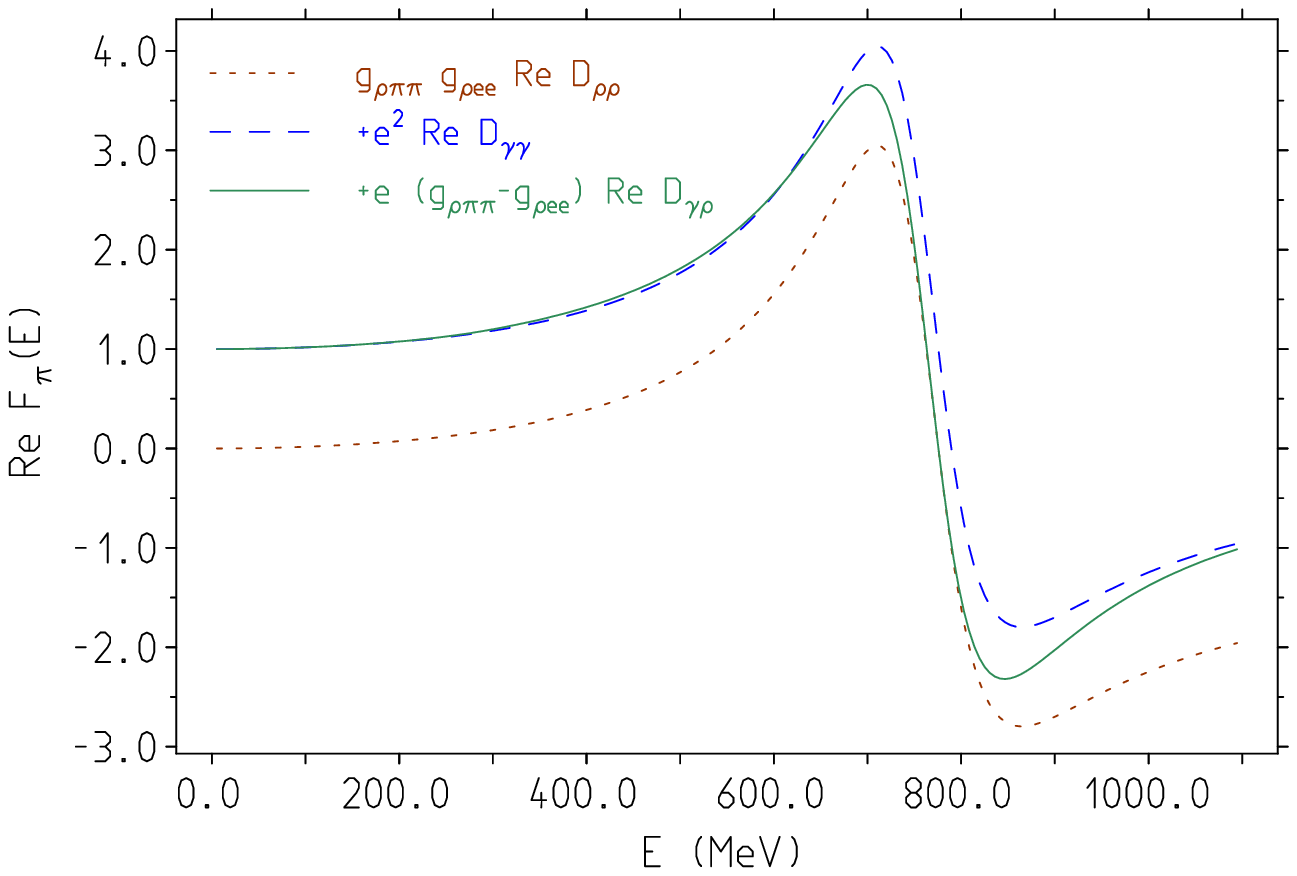}
\includegraphics[height=5cm]{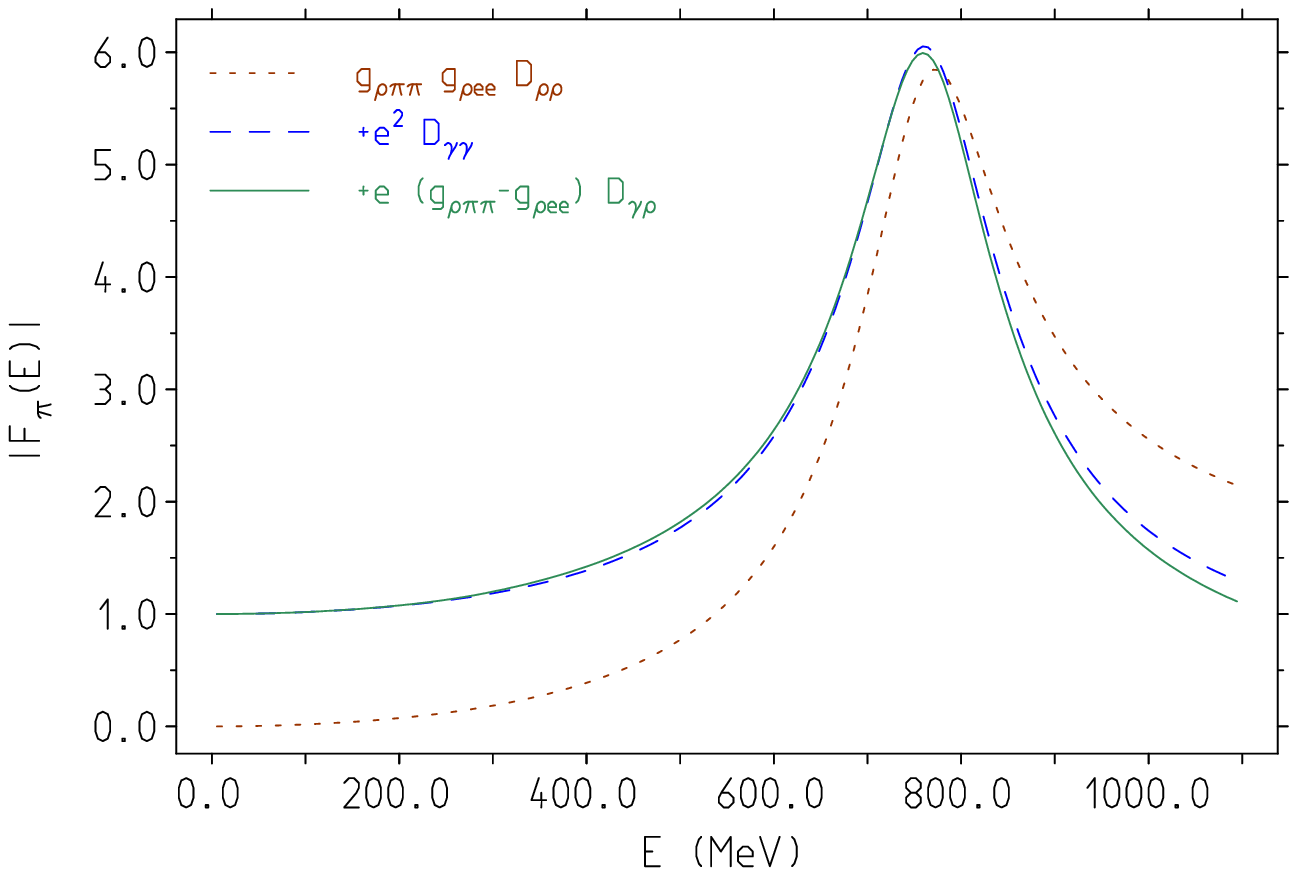}
\caption{The real parts and moduli of the three terms of
(\ref{fpidecomp}), individual and added up.}
\label{fig:terms}
\end{figure}

Real parts and moduli of the individual terms normalized to the sQED
photon exchange term are displayed in Fig.~\ref{fig:terms}.

\begin{figure}
\centering
\includegraphics[height=7cm]{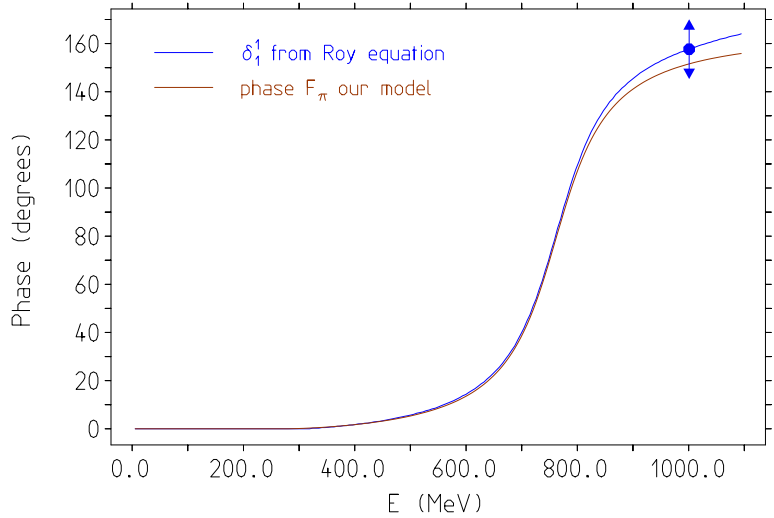}
\caption{The phase of $F_\pi(E)$ as a function of the c.m. energy
E. We compare the result of the elaborate Roy equation analysis of Ref.~\cite{LeCo02}
with the one due to the sQED pion-loop. The solution of the Roy equation
depends on the normalization at a high energy point (typically 1
GeV). In our calculation we could adjust it by varying the coupling $g_{\rho\pi\pi}$.}
\label{fig:fpiphase}
\end{figure}

An improved theory of the pion form factor has been developed in~\cite{LeCo02}. 
One of the key ingredients in this
approach is the strong interaction phase shift $\delta^1_1(s)$ of
$\pi\pi$ (re)scattering in the final state. In Fig.~\ref{fig:fpiphase}
we compare the phase of $F_\pi(s)$ in our model with the one obtained
by solving the Roy equation with $\pi\pi$-scattering data as input.
We notice that the agreement is surprisingly good up to about 1 GeV.
It is not difficult to replace our phase by the more precise exact one.

We note that the precise $s$-dependence of the effective
$\rho$-width is obtained by evaluating the imaginary part of the $\rho$
self-energy:
\ba
\Impa \Pi_{\rho
\rho}=\frac{g_{\rho\pi\pi}^2}{48\,\pi}\,\beta_\pi^3\,s\equiv M_\rho\,\Gamma_\rho(s)\cs
\ea
which yields
\ba
\Gamma_\rho(s)/M_\rho=\frac{g_{\rho\pi\pi}^2}{48\,\pi}\,\beta_\pi^3\,\frac{s}{\mr}
\semis \Gamma_\rho/M_\rho=\frac{g_{\rho\pi\pi}^2}{48\,\pi}\,\beta_\rho^3\epo
\ea
In our model, in the given approximation, the  on $\rho$-mass-shell
form factor reads
\ba
F_\pi(M_\rho^2)=1- \I\, \frac{g_{\rho ee}g_{\rho\pi\pi}}{e^2}\,\frac{M_\rho}{\Gamma_\rho}\cs
\ea
and the square modulus may be written as
\ba
|F_\pi(M_\rho^2)|^2=1+ \frac{36}{\alpha^2}\frac{\Gamma_{ee}}{\beta_\rho^3\,\Gamma_\rho}\cs
\ea
with
\ba
\Gamma_{\rho ee}=\frac{1}{3}\, \frac{g^2_{\rho ee}}{4\pi}\, M_\rho
\; \mathrm{ \ or \ }\;\;g_{\rho ee}=\sqrt{12\pi\,\Gamma_{\rho
ee}/M_\rho} \epo
\ea
It is interesting to note that the GS formula (\ref{GSfin}) does not involve
$\Gamma_{\rho ee}$ in any direct way, since the normalization is fixed
by applying an overall factor $1+d\,\Gamma_\rho/M_\rho\equiv 1-\Pi^\mathrm{ren}_{\rho \rho}(0)/M_\rho^2$
to enforce $F_\pi(0)=1$. The leptonic width is then given by
\ba
\Gamma^{\mathrm{GS}}_{\rho ee}=\frac{2\alpha^2\,\beta^3_\rho M_\rho^2}{9\,\Gamma_\rho}
\,\left(1+d\,\Gamma_\rho/M_\rho\right)^2 \ep
\ea
In the CMD-2 fit $1+d\,\Gamma_\rho/M_\rho \simeq 1.089$.

\begin{figure}[t]
\centering
\includegraphics[height=8cm]{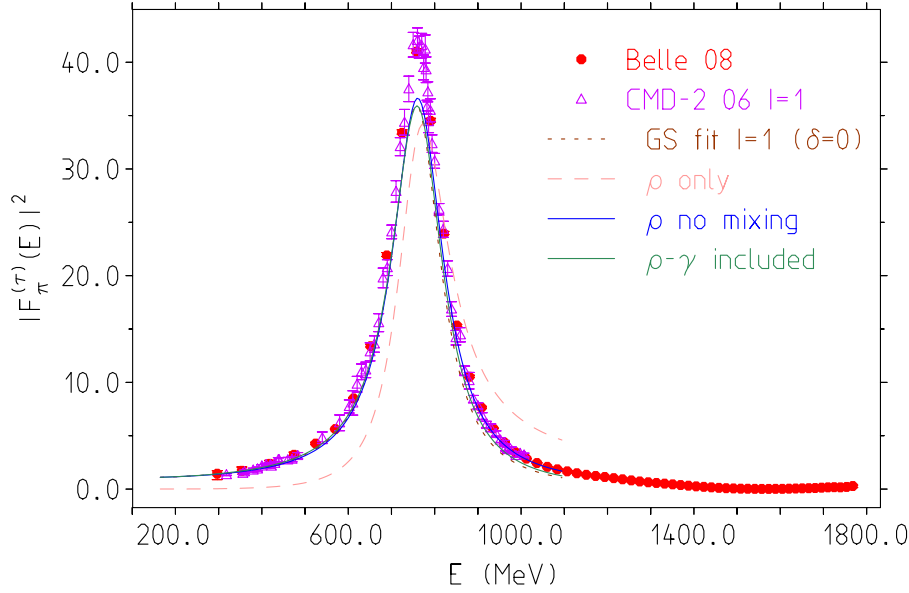}
\caption{Effect of $\gamma$ - $\rho$ mixing based on the simple EFT model (\ref{ModelSimple}).
Parameters: $M_\rho=775.5~\mv$, $\Gamma_\rho=143.85~\mv$,
$\cB[(\rho\to ee)/(\rho\to \pi\pi)]=4.67\power{-5}$, $e=0.302822$,
$g_{\rho \pi\pi}=5.92$, $g_{\rho ee}=0.01826$.
The crucial point is the difference between ``$\rho$ no mixing'' [1st
plus 3rd term of (\ref{fpidecomp})] and
``$\rho-\gamma$ mixing included'' [all terms of (\ref{fpidecomp})]. The interference with the mixing
term lowers the form factor above the $\sqrt{s}\sim M_\rho$, an effect
not present in the charged (pure $I=1$) channel. For comparison the GS
fit with switched off $\rho'$ and $\rho''$ and the ``$\rho$ only''
[3rd term of (\ref{fpidecomp}) only]
are shown.}.
\label{fig:BWgamma}
\end{figure}

The result for $|F_\pi(s)|^2$ (using mass and width as
before) is displayed in Fig.~\ref{fig:BWgamma}.
We compare the results obtained when $\rho$ - $\gamma$
mixing is properly taken into account with the one obtained by
ignoring mixing and with a GS fit with just the $\rho$ taken into account.
At first look, the results agree fairly well but do not fit the Belle data
as expected if we do not include the higher resonances.

\begin{figure}[t]
\centering
\includegraphics[height=6cm]{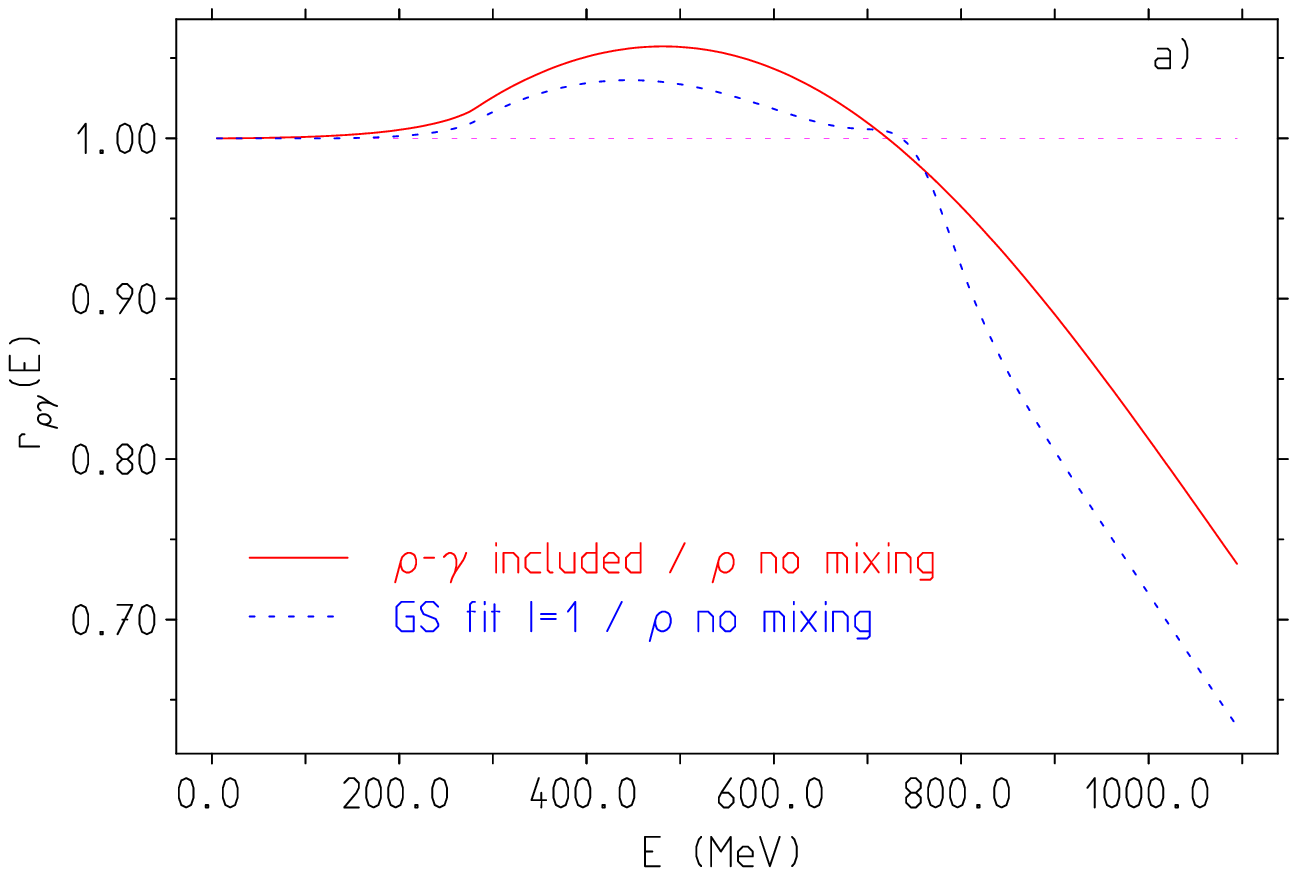}
\includegraphics[height=6cm]{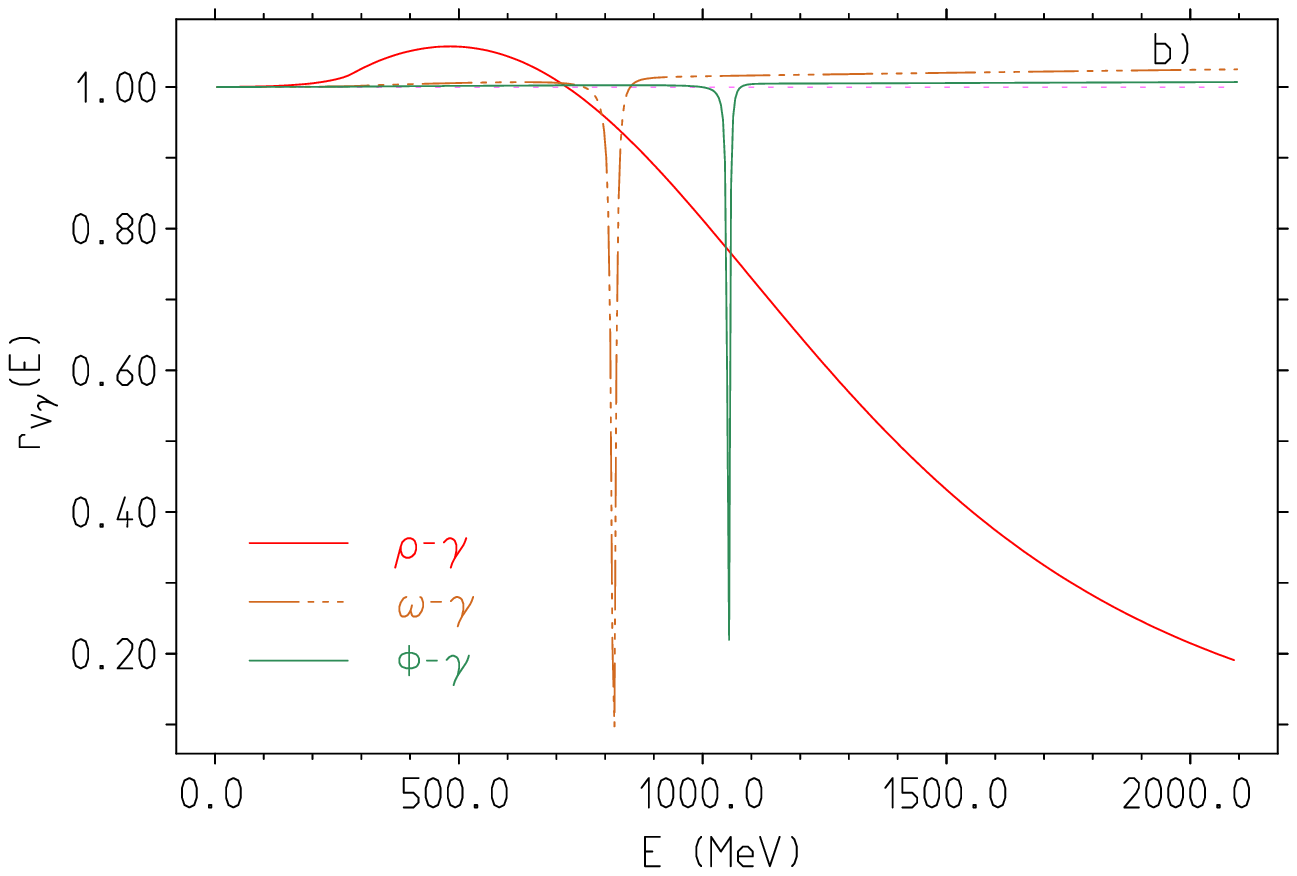}
\caption{a) Ratio of the full $|F_\pi(E)|^2$
in units of the same quantity omitting the mixing term together with
a standard GS fit with PDG parameters. b)  
The same mechanism scaled up by the branching fraction $\Gamma_V/\Gamma(V\to\pi\pi)$
for $V=\omega$ and $\phi$. In the $\pi\pi$ channel the
effects for resonaces $V\neq \rho$ are tiny if not very close to resonance.}
\label{fig:mixingcorr}
\end{figure}

A detailed comparison, in terms of the ratio
\ba
r_{\rho\gamma}(s)\equiv \frac{|F_\pi(s)|^2}{|F_\pi(s)|^2_{D_{\gamma \rho}=0}}\cs
\label{rrg}
\ea
shown in Fig.~\ref{fig:mixingcorr}, however reveals substantial
differences and proves the relevance of the mixing. We also plotted
the same ratio for the I=1 part of the GS fit, which exhibits a
similar behavior as the true $F_\pi(s)$. This is not really surprising
as one fits the same data just in a different way, i.e. with different
parameters for the $\rho$. This mixing affects, however, the
relationship to the $\tau$ channel, which does not exhibit this
effect. Of course at higher energies, not to far above the $\rho$, it
is not known whether the simple EFT model can be trusted. Note that
dropping the $\Pi^2_{\gamma \rho}$ terms (approximation indicated in
(\ref{invpropterms})) in the Dyson resummed propagators does not
affect the result. We have checked that $\omega-\gamma$, $\phi-\gamma$
or $\Upsilon(4S)-\gamma$ mixing effects are tiny away from the
resonances and thus should not affect the interpretation of radiative
return spectra as measured at KLOE and BaBar.

\begin{figure}[t]
\centering
\includegraphics[height=5cm]{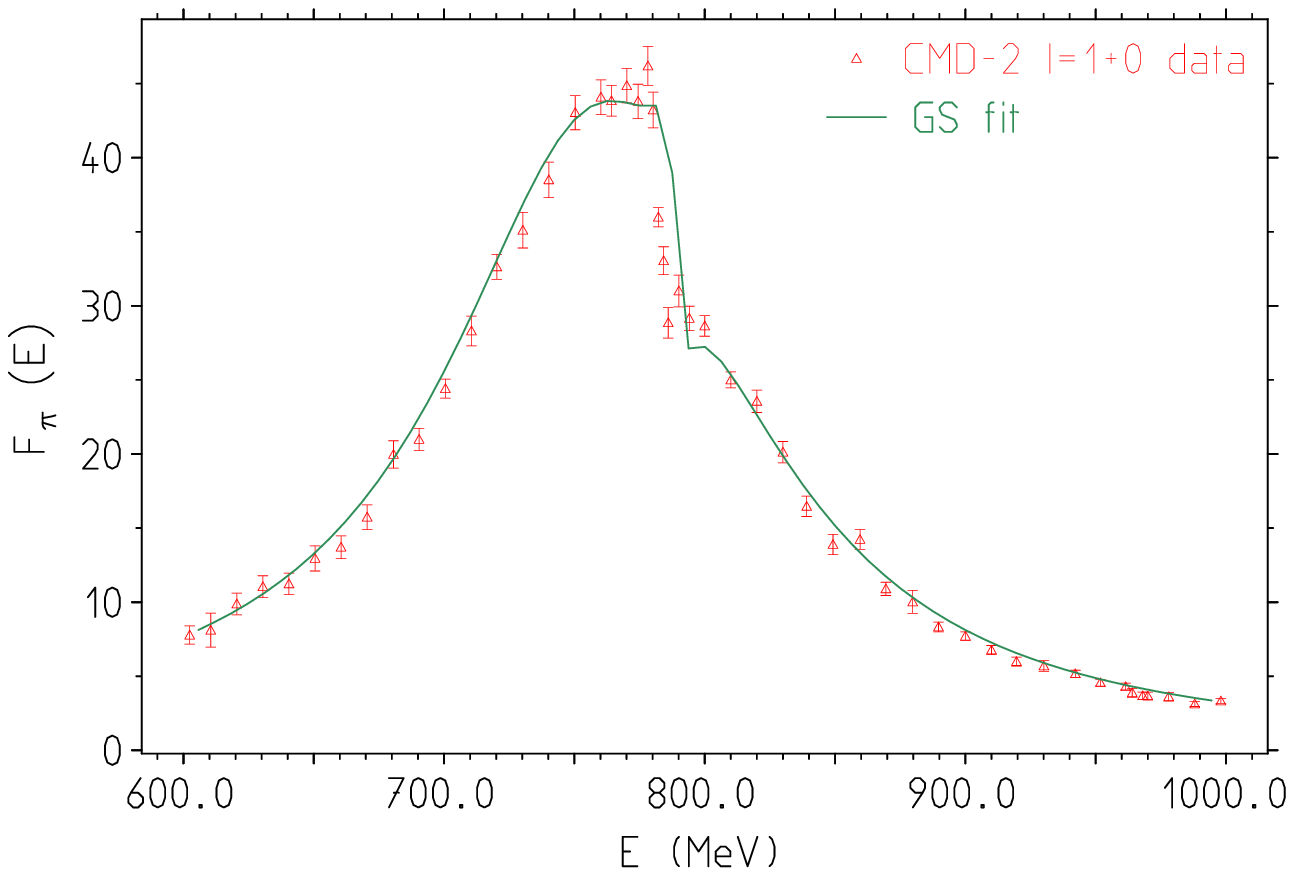}
\includegraphics[height=5cm]{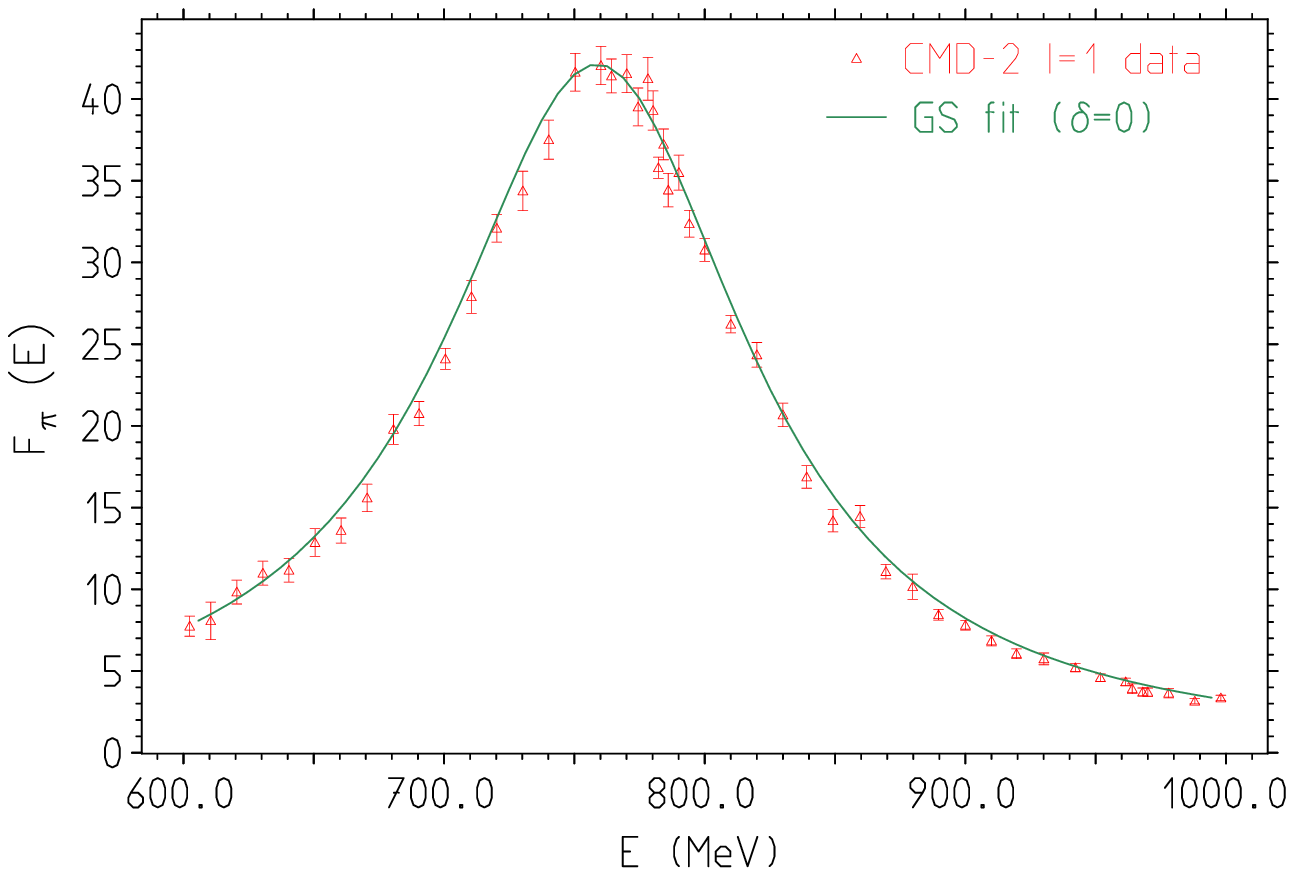}
\caption[]{CMD-2 data for $|F_\pi|^2$ in $\rho-\omega$ region together with
Gounaris-Sakurai fit. Left before subtraction right after subtraction of the $\omega$.}
\label{fig:rhoomegasubtr}
\end{figure}

We have to compare the model with the $I=1$ part of the $\epm$-data.
To this end we may take the CMD-2 fit of the CMD-2 data~\cite{CMD206} and set the
mixing parameter $\delta=0$ as illustrated in
Fig.~\ref{fig:rhoomegasubtr}. In this way we obtain the isovector part of
the square of the pion form factor $|F^{(e)}_\pi[I=1](s)|^2$.

In order to compare $F^{(\tau)}_\pi(s)$ extracted from $\tau$-decay
spectra with $F^{(e)}_\pi(s)$ measured in $\epm$-annihilation we have
to apply isospin breaking corrections as investigated
in~\cite{CEN} and~\cite{FloresTlalpa:2005fz} (see also~\cite{DEHZ03,GJ03,Davier:2009ag}):

\bit
\item Mass shift: the Cottingham formula, which allows us for
a rather precise calculation of the electromagnetic pion mass shift
$\delta m_\pi = m_{\pi^\pm}-m_{\pi^0}\simeq 4.6~\mv$, suggest the
relation $\Delta m_\pi^2 \simeq \Delta M_\rho^2$, which yields a shift
between changed an neutral $\rho$ by $\delta
M_\rho=M_{\rho^\pm}-M_{\rho^0}\simeq\ha\,\frac{\Delta
m_\pi^2}{M_{\rho^0}}=0.814~\mv$.

\item Width shift: The kinematic shift form pion and rho mass differences
is $\delta \Gamma_\rho=\Gamma_{\rho^\pm}-\Gamma_{\rho^0}=\frac{g^2_{\rho
\pi\pi}}{48\pi}\,(\beta^3_{\rho^\pm}\,M_{\rho^\pm}-\beta^3_{\rho^0}\,M_{\rho^0})
=1.300~\mv$.

\item Pion velocities are $\beta_{\pi,0}=\sqrt{1-\frac{(2m_{\pi^\pm})^2}{s}}$ and
$\beta_{\pi,\pm}=\sqrt{1-\frac{(m_{\pi^\pm}+m_{\pi^0})^2}{s}
-\frac{(m_{\pi^\pm})-m_{\pi^0}^2}{s}}$ for neutral and charged
channel. respectively. Their on-resonance values read
$\beta_{\rho^\pm}=\beta_{\pi,\pm}(s=M^2_{\rho^\pm})$ and
$\beta_{\rho^0}=\beta_{\pi,0}(s=M^2_{\rho^0})$.

\item In the charged
channel ($\tau$-decay) the appropriate phase-space for the
$\pi^\pm\pi^0$ system, replacing the $\pi^+\pi^-$ one, has to be
considered. For the energy dependent width one has
$\Gamma_{\rho^\pm}(s)=\Gamma_{\rho^\pm}\,\frac{\beta^3_{\pi,\pm}}{\beta^3_{\rho^\pm}}
\frac{s}{M^2_{\rho^\pm}}$.
\eit
We have made use of the fact that the strong coupling factor
$\frac{g^2_{\rho
\pi\pi}}{48\pi}=\frac{\Gamma_{\rho^0}}{M_{\rho^0}\,\beta^3_{\rho^0}}=
\frac{\Gamma_{\rho^\pm}}{M_{\rho^\pm}\,\beta^3_{\rho^\pm}}$ is charge independent.
Note that all of these corrections represent corrections in $F_0(s)/F_-(s)$
the ratio between neutral and charged channel $F_\pi$'s. 
\bit
\item Electromagnetic corrections $G_{\mathrm{EM}}(s)$ as calculated in~\cite{CEN,FloresTlalpa:2005fz}.
Specifically, we will apply the correction given by~\cite{CEN}, since
the ones given in~\cite{FloresTlalpa:2005fz} differ quite a lot for
reasons we have not yet understood. It does not affect the main
conclusion of our analysis.
\eit
In total a correction\footnote{If we would not include the $\rho-\gamma$
mixing in $F_0(s)$ the correction formula would read
$$v_0(s)=r_{\rho\gamma}(s)\,R_{\rm IB}(s)\,v_-(s)\epo$$}  
\ba
v_0(s)=R_{\rm IB}(s)\,v_-(s)\semis R_{\mathrm{IB}}(s)=\frac{1}{G_{\rm EM}(s)} 
\frac{\beta_0^3(s)}{\beta_-^3(s)}\left|\frac{F_0(s)}{F_-(s)}\right|^2
\ea
has to be applied in the relation between the spectral functions.
Final state radiation correction FSR(s) and vacuum
polarization effects we have been subtracted from all $\epm$-data.
\begin{figure}[t]
\centering
\includegraphics[height=5cm]{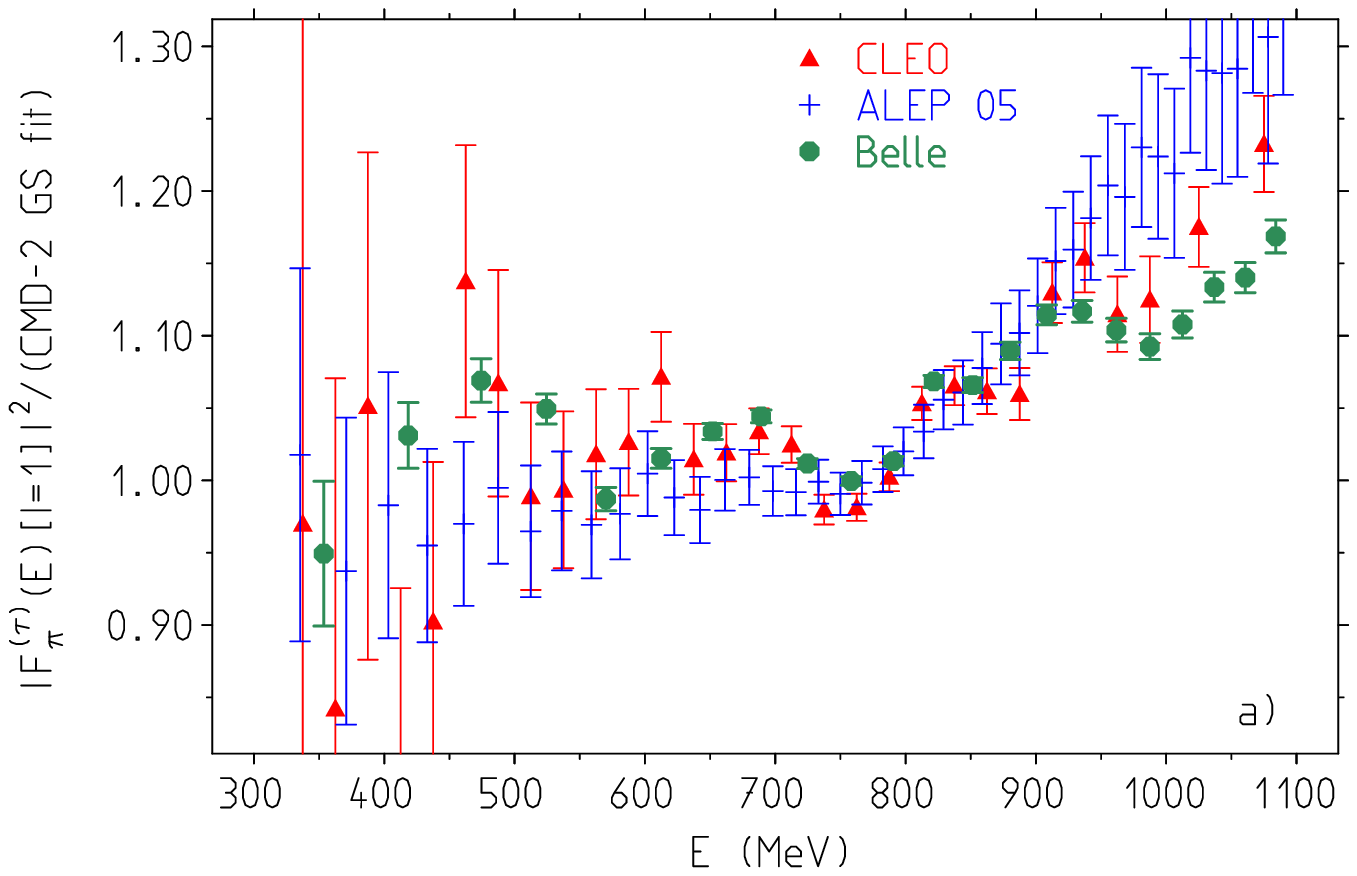}
\includegraphics[height=5cm]{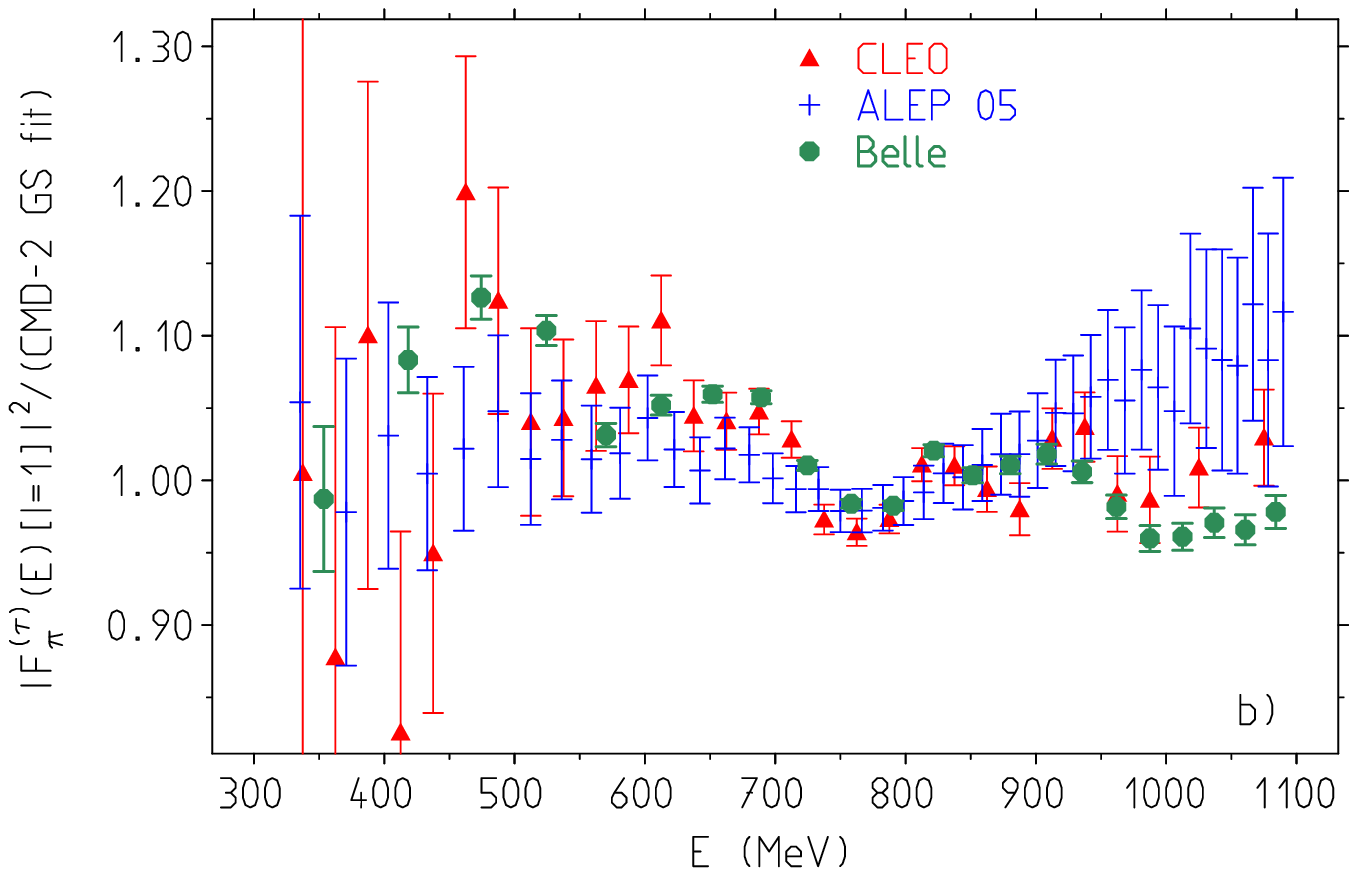}
\includegraphics[height=5cm]{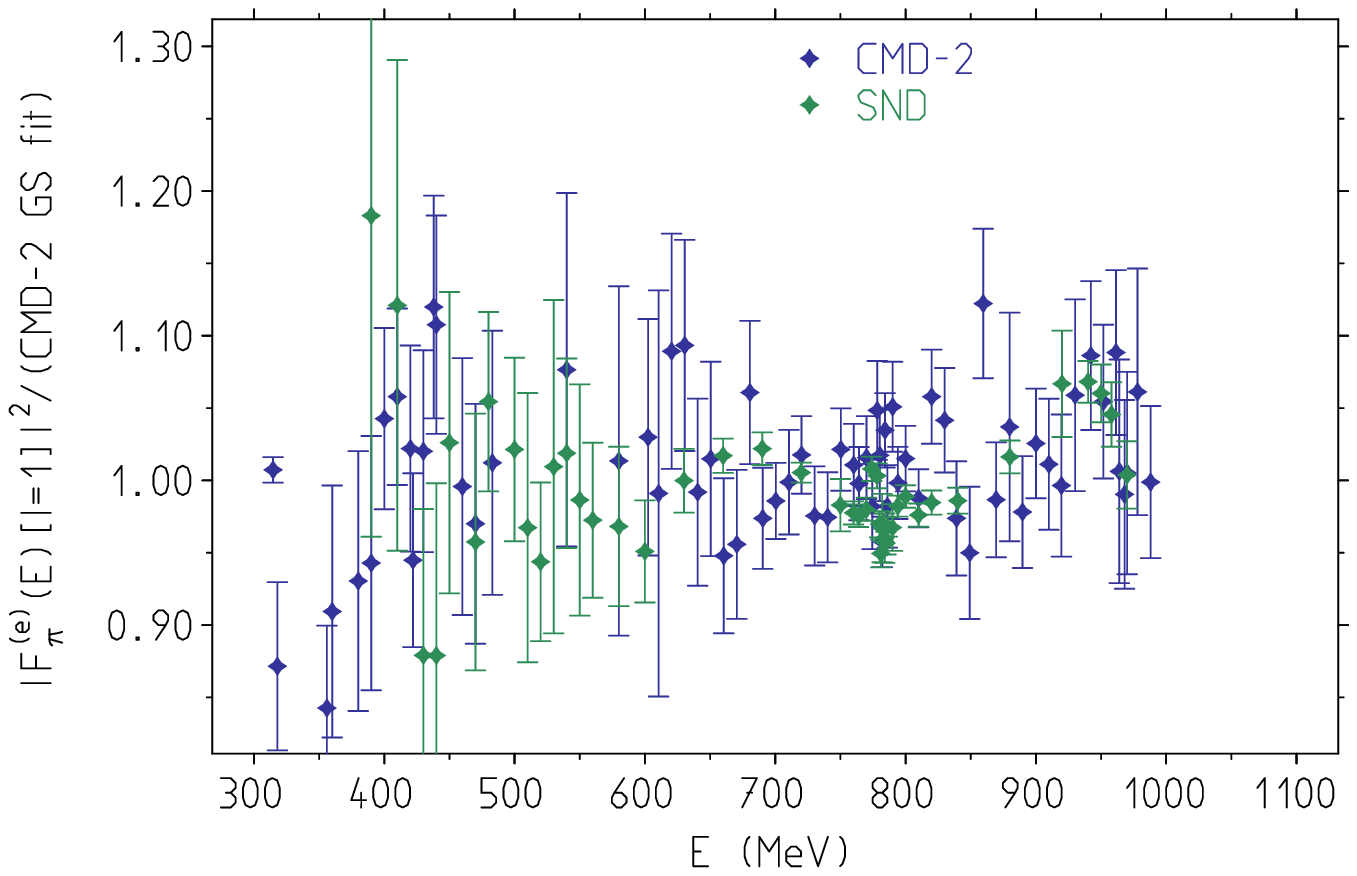}
\includegraphics[height=5cm]{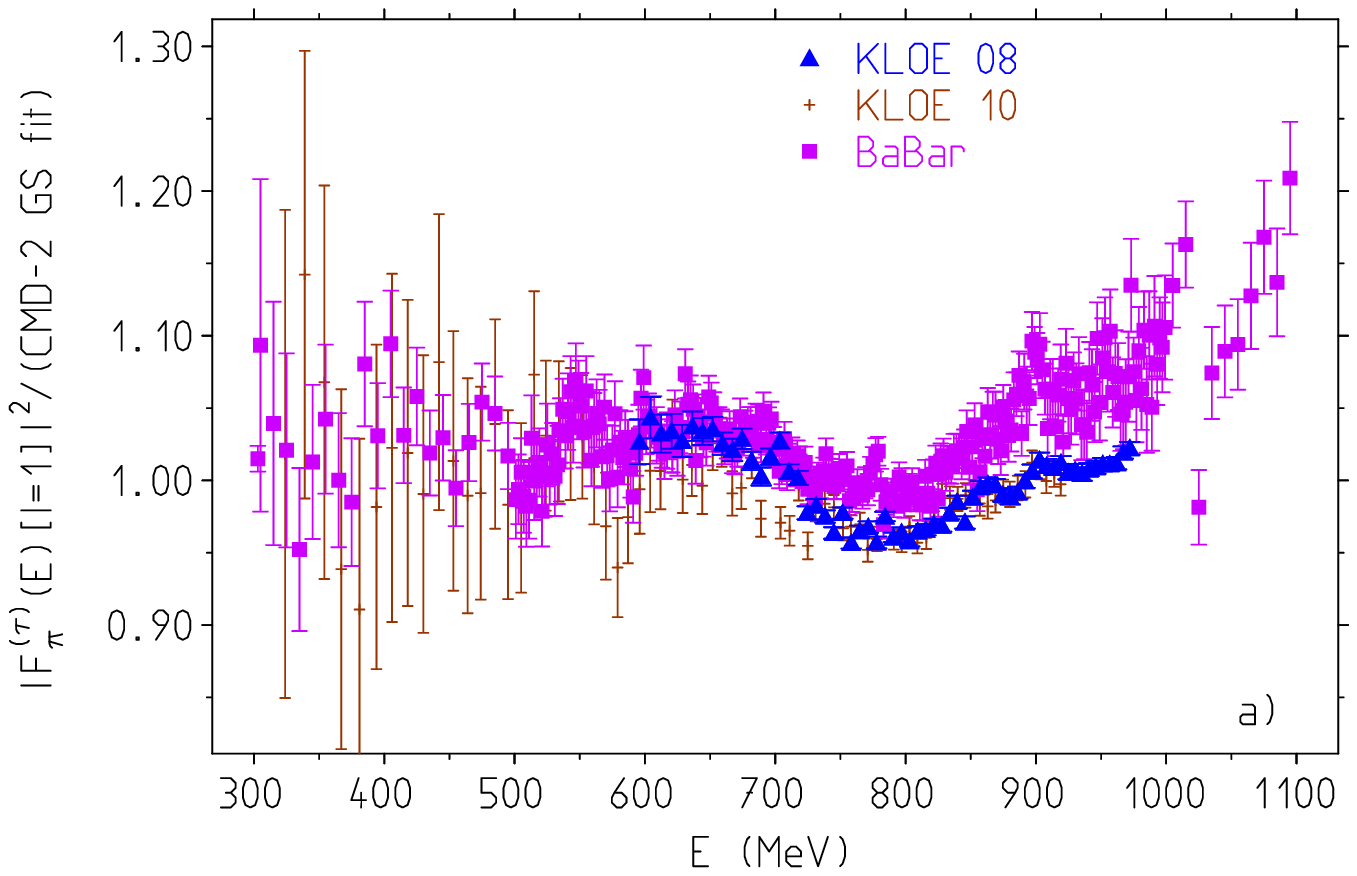}
\caption{$|F_\pi(E)|^2$ ratio $\tau$ vs. $\epm$ I=1 (CMD-2 GS fit): a) uncorrected for $\rho -
\gamma$ mixing, and b) the same after correcting for it. The
correction factor is given by the solid curve of
Fig.~\ref{fig:mixingcorr} a). Lower panel: $\epm$ energy scan data
[left] and $\epm$ radiative return data [right]. The GS fit chosen as
a reference is represented by the full line in the right
Fig.~\ref{fig:rhoomegasubtr}. The I=1 part for the $\epm$ sets is
obtained by subtracting the difference of the two curves shown in
Fig.~\ref{fig:rhoomegasubtr}. The choice of the particular reference
is of course ambiguous.}
\label{fig:dataratios}
\end{figure}

In Fig.~\ref{fig:dataratios} we illustrate the consequence of $\rho
-\gamma$ mixing. After applying the correction (for our set of
parameters, which is not far from standard GS fit parameters) the
consistency of $\tau$ and $\epm$ data is dramatically
improved. However, substantial differences of different measurements
remain as a problem.

How does the new correction affect the evaluation of the hadronic
contribution to the anomalous magnetic moment of the muon? To
lowest order in terms of $\epm$-data, represented by $R(s)$, we have
\ba
a_\mu^{\mathrm{had,LO}}(\pi\pi)=
\frac{\alpha^2}{3\pi^2}\,\int_{4m_\pi^2}^{\infty}\,\D s\,R^{(0)}_{\pi\pi}(s)\frac{K(s)}{s}\cs
\ea
with the well-known kernel $K(s)$ and $$R^{(0)}_{\pi\pi}(s)=(3 s
\sigma_{\pi\pi})/4\pi \alpha^2(s))=3 v_0(s)\epo$$ Note that the
$\rho-\gamma$ interference is included in the measured $\epm$-data,
and so is its contribution to $\amuh$. In fact $\amuh$ is intrinsic an
$\epm$-based ``observable'' (neutral current channel). If we want to
use the $\tau$ spectral function\footnote{In the exact $SU(2)$
(isospin) limit we would have the CVC
relation $v_-(s)=v_0(s)$ and the $\tau$ data would give the same
contribution as the $\epm$ data.} isospin breaking corrections to the
$\tau$ data must be applied: traditionally $v_-(s) \to v_0(s)=R_{\rm
IB}(s)\,v_-(s)$. For our comparison we switch off the I=0
($\rho-\omega$ mixing part) part from the $\epm$ data, which we do not
include in $R_{\rm IB}$. In addition we have to account for the
missing $\rho-\gamma$ mixing in the $\tau$ spectra. The results for
the I=1 part of $\amuh$ is given in the Tab.~\ref{tab:amuhad}.
\begin{table}[t]
\centering
\caption{Isovector (I=1) contribution to $\amuh \times 10^{10}$ from the range [0.592 - 0.975]
GeV from selected experiments. First entry: results from $\tau$-data after standard isospin
breaking (IB) corrections. Second entry: results from $\tau$-data after
applying in addition the $\rho -\gamma$ mixing corrections $r_{\rho\gamma}(s)$, with
fitted values for $M_\rho,\Gamma_\rho$ and $\Gamma_{\rho ee}$
[$M_\rho=775.65~\mv,\Gamma_\rho=149.99~\mv,
\cB[(\rho\to ee)/(\rho\to \pi\pi)]=4.10\power{-5}$]. Third
entry: as second one, however using PDG values for the $\rho$
parameters [$M_\rho=775.5~\mv,\Gamma_\rho=143.85~\mv,
\cB[(\rho\to ee)/(\rho\to \pi\pi)]=4.67\power{-5}$]  (for illustration of sensitivity to
the choice of parameters). For the $\rho-\omega$ mixing we subtracted
$2.67\power{-10}$. Errors are statistical, systematic, isospin breaking
and $\rho-\gamma$ mixing, assuming a 10\% uncertainty for the
latter. Final state radiation is not included. }
\label{tab:amuhad}
\begin{tabular}{c|c|c|c}
\hline
 Data &~~  standard IB corrections ~~&~~ incl. $\rho-\gamma$ mixing~~ &~~ same
 using GS fit parameters for the $\rho$~~  \\
\hline
ALEPH 1997~\cite{ALEPH97} & 390.75(2.69)(1.97)(1.45) & 385.63(2.65)(1.94)(1.43)(0.50) & $-$\\
ALEPH 2005~\cite{ALEPH05} & 388.74(4.05)(2.10)(1.45) & 383.54(4.00)(2.07)(1.43)(0.50) & 385.05(4.01)(2.08)(1.44)(0.36)\\
OPAL~ 1999~\cite{OPAL99} & 380.25(7.36)(5.13)(1.45) & 375.39(7.27)(5.06)(1.43)(0.50) & $-$\\
CLEO~ 2000~\cite{CLEO00} & 391.59(4.16)(6.81)(1.45) & 386.61(4.11)(6.72)(1.43)(0.50) & 388.10(4.12)(6.75)(1.44)(0.36) \\
BELLE 2008~\cite{Belle08} & 394.67(0.53)(3.66)(1.45) & 389.62(0.53)(3.66)(1.43)(0.50) & 391.06(0.52)(3.63)(1.44)(0.36)\\
\hline
average & 391.06(1.42)(1.47)(1.45) & 385.96(1.40)(1.45)(1.43)(0.50) & \\
\hline
CMD-2 2006~\cite{CMD206} &     &  386.34(2.26)(2.65) & \\
SND~~ 2006~\cite{SND06} &     &  383.99(1.40)(4.99) & \\
KLOE~ 2008~\cite{KLOE08} &     &  380.24(0.34)(3.27) & \\
KLOE~ 2010~\cite{KLOE10} &     &  377.35(0.71)(3.50) & \\
BABAR 2009~\cite{BABAR09} &     &  389.35(0.37)(2.00) & \\
\hline
average &    &  385.12(0.87)(2.18) & \\
\hline
all $\epm$ data &  & 385.21(0.18)(1.54) & \\
$\epm$ + $\tau$ &  & 385.42 (0.53)(1.21) & \\
\hline
\end{tabular}
\end{table}
Remarkably, the contributions to $\amuh$ from the $\epm$-data on the
one hand and from the $\tau$ data on the other hand agree surprisingly
well after including the mixing. The $\rho-\gamma$ mixing correction
has been applied for the $\rho$ part only ($\rho'$ and $\rho''$
subtracted and added back with the help of the GS fit from Belle). We
do not attempt here a complete analysis of all $\tau$ data. The
$\rho-\gamma$ mixing contribution, which implies a moderate positive
interference below the $\rho$ resonance and a stronger negative one
above the $\rho$ resonance, shift the $\tau$ data to lie perfectly
within the ballpark of the $\epm$ data\footnote{Note that the moderate
shifts of rho masses and widths~\cite{GJ03} which diminish somewhat
the discrepancy are included in the IB corrections, as detailed
above.}.

Another important quantity which can be directly measured is the
branching fraction $B^{\mathrm{CVC}}_{\pi\pi^0}=
\Gamma (\tau \to \nu_\tau \pi\pi^0)/\Gamma _\tau$. This
``$\tau$-observable'' can be evaluated in terms of the I=1 part of the
$\eepp$ cross section, after taking into account the IB correction
$v_0(s) \to v_-(s)= v_0(s)/R_{\rm IB}(s)$,
\ba
B^{\mathrm{CVC}}_{\pi\pi^0}= \frac{2S_\mathrm{EW}B_e |V_{ud}|^2}{m_\tau^2}\,
\int_{4m_\pi^2}^{m_\tau^2}\,\D s \,R^{(0)}_{\pi^+\pi^-}(s)\,\left(1-\frac{2}{m_\tau^2}\right)^2
\left(1+\frac{2s}{m_\tau^2}\right)\,\frac{1}{R_{\mathrm{IB}}(s)}\cs
\ea
where here we also have to ``undo'' the
$\rho-\gamma$ mixing which is absent in the charged isovector
channel. Results are given in Tab.~\ref{tab:BCVC}. 
The shift by the $\rho-\gamma$ mixing is +0.62 \% to which we assign an
error of 10 \% \footnote{Averages given 
in~\cite{Davier:2009ag,Hoecker:2010qn} are 25.42 $\pm$ 0.10 \% for $\tau$ and 24.78 $\pm$ 0.28 \% 
for $\epm$ + CVC. Adding the new correction we get 25.40 $\pm$ 0.28 $\pm$ 0.06 \%
for the latter, in perfect agreement with the $\tau$ result. The BaBar
di-pion spectral function, not included in the previous numbers,
yields 25.15 $\pm$ 0.18 $\pm$ 0.22 \% or 25.77 $\pm$ 0.18 $\pm$ 0.28
including the new correction. Results differ slightly from ours
because we apply slightly different IB corrections.}.
\begin{table}[t]
\centering
\caption{Calculated branching fractions in \% from selected experiments.
Experimental data completed down to threshold and up to $m_\tau$ by
corresponding world averages where necessary. CMD-2 is a combination
of 2003 and 2006 data. The experimental world average of direct
branching fractions is
$B^{\mathrm{CVC}}_{\pi\pi^0}$ = 25.51 $\pm$ 0.09 \% ~\cite{PDG10}.}
\label{tab:BCVC}

\begin{tabular}{lllc|c||lllc|c}
\hline
 \multicolumn{3}{c}{ $\tau$ experiments~~~} ~~& &~~~ $B_{\pi\pi^0}$[\%]&
\multicolumn{3}{c}{ $~~\epm$ experiments~~~} ~~& &~~~ $B^{\mathrm{CVC}}_{\pi\pi^0}$[\%]\\
\hline
&ALEPH& 1997~~~ &&~~ 25.27 $\pm$ 0.17 $\pm$ 0.13 ~~& &~~ CMD-2& 2006~~~ &&~~ 25.40 $\pm$
0.21 $\pm$ 0.28 ~~\\
&ALEPH& 2005 && 25.40 $\pm$ 0.10 $\pm$ 0.09 &	  &~~  SND & 2006 && 25.09
$\pm$ 0.30 $\pm$ 0.28\\
&OPAL& 1999 && 25.17 $\pm$ 0.17 $\pm$ 0.29&	  &~~ KLOE& 2008 && 24.82
$\pm$ 0.29 $\pm$ 0.28\\
&CLEO& 2000 && 25.28 $\pm$ 0.12 $\pm$ 0.42& 	  &~~ KLOE& 2010 &&  24.65
$\pm$ 0.29 $\pm$ 0.28\\
&Belle& 2008 && 25.40 $\pm$ 0.01 $\pm$ 0.39&	  &~~ BaBar& 2009 && 25.45
$\pm$ 0.18 $\pm$ 0.28\\
\hline
& \multicolumn{2}{c}{combined} && 25.34 $\pm$ 0.06 $\pm$ 0.08 & & \multicolumn{2}{c}{combined} && 25.20 
$\pm$ 0.17 $\pm$ 0.28 \\
\hline
\end{tabular}
\end{table}
For the direct $\tau$ branching fractions the first error is
statistical the second systematic. For $\epm$+CVC the first error is
experimental the second error includes uncertainties of the IB
correction +0.06 from the new mixing effect. Remaining problems seem
to be experimental, there are significant differences in the spectral
functions from different experiments (see Fig.~\ref{fig:dataratios}).

%
%\begin{figure}[t]
%\centering
%\includegraphics[height=7cm]{BCVCtauvsepem}
%\caption{Branching fractions $B(\tau \to \pi\pi^0 \nu_\tau)$ obtained
%from $\tau$ spectra and from $\epm$-data via CVC after appying 
%IB corrections. The substantial difference seen when $\rho-\gamma$
%mixing is not taken care off essentally disappears
%after including the correction $r_{\rho\gamma}(s)$ (\ref{rrg}). 
%The vertical bar represents the experimental world average.}
%\label{fig:BCVC} 
%\end{figure}
%
\section{$\rho-\omega$ mixing}
In order to include the I=0 contribution form $\omega \to \pi^+\pi^-$  we
need to consider the corresponding symmetric ($\gamma$, $\rho$, $\omega$) 3$\times$3
matrix propagator, with new entries $\Pi_{\gamma \omega}(q^2)$,  $\Pi_{\rho \omega}(q^2)$
and $q^2-M_\omega^2+\Pi_{\omega \omega}(q^2)$, supplementing the
inverse propagator matrix (\ref{invprop22}) by a 3rd
row/column. Treating all off-diagonal elements as perturbations (after
diagonalization) to linear order the new elements in the propagator read:
\ba
D_{\gamma \omega} &\simeq& \frac{-\Pi_{\gamma \omega}(q^2)}{(q^2+\Pi_{\gamma \gamma}(q^2))
\,(q^2-M_\omega^2+\Pi_{\omega\omega}(q^2))}  \crn
D_{\rho \omega} &\simeq& \frac{-\Pi_{\rho \omega}(q^2)}{(q^2-\mr +\Pi_{\rho\rho}(q^2))
\,(q^2-M_\omega^2+\Pi_{\omega \omega}(q^2))}  \crn
D_{\omega\omega} &\simeq& \frac{1}{q^2-M_\omega^2 +\Pi_{\omega\omega}(q^2)}\;.
\ea
The self-energies again are the renormalized ones and in the two pion
channel $\eepp$ given up to different coupling factors by the
same self-energy functions as in the $\gamma -\rho$ sector. Thus, the
bare self-energy functions read
\ba
\Pi_{\gamma \omega}&=&\frac{e g_{\omega \pi\pi}}{48 \pi^2}\,f(q^2)\comas
\Pi_{\rho \omega}=\frac{g_{\rho\pi\pi} g_{\omega\pi\pi}}{48 \pi^2}\,f(q^2)\;\: \mathrm{ \ and \ }\;\;
\Pi_{\omega \omega}=\frac{g^2_{\omega \pi\pi}}{48 \pi^2}\,f(q^2)\comas
\ea
and they are renormalized analogous to (\ref{grren},\ref{rrren})
subtracted at the $\omega$ mass shell. The $\rho - \omega$ mixing term is
special here because if we diagonalize it on the $\rho$ mass shell
the matrix is no longer diagonal at the $\omega$-resonance, where
\ba
\Pi^\mathrm{ren}_{\rho \omega}(q^2)&=&  \Pi_{\rho \omega}(q^2)
-\frac{q^2}{M_\rho^2}\,\Repa\Pi_{\rho \omega}(M_\rho^2) \stackrel{q^2=\mo}{\to}
\Pi_{\rho \omega}(\mo)
-\frac{\mo}{M_\rho^2}\,\Repa\Pi_{\rho \omega}(M_\rho^2)\neq 0\cs
\ea
and which yields the leading I=0 contribution to the pion form
factor\footnote{Typically, $\Pi^\mathrm{ren}_{\gamma \rho}(\mo)=
\frac{e g_{\rho \pi\pi}}{48 \pi^2}\,\mo\,\left(h(\mo)-\Repa h(\mr)\right)$ and
$D_{\gamma \rho}(\mo)=-\frac{e g_{\rho \pi\pi}}{48 \pi^2}
\frac{\left(h(\mo)-\Repa h(\mr)\right)}{\mo-\mr+\I\,M_\rho\,\Gamma_\rho}$.
Similarly, $D_{\rho \rho}(\mo)=\frac{1}{\mo-\mr +\I \,M_\rho
\Gamma_\rho}$ taking
$\Gamma_\rho(\mo) \sim \Gamma_\rho\epo$ }.
The $\omega$ induced terms contribute to the pion form factor
\ba
\Delta F^{(\omega)}_\pi(s)= \left[e\,\left(g_{\omega\pi\pi}-g_{\omega e
e}\right)\,D_{\gamma \omega}-(g_{\rho e e} g_{\omega \pi\pi}+g_{\omega
ee}g_{\rho
\pi\pi})\,D_{\rho \omega} \right]/\left[e^2\,D_{\gamma\gamma}\right]\cs
\ea
which adds to (\ref{fpidecomp}). The direct $~~\epm \to \omega \to
\pi\pi~~$ term given by $- g_{\omega \pi\pi}g_{\omega e e}\,D_{\omega
\omega}$ by convention is taken into
account as part of the complete $\omega$-resonance contribution.

\begin{figure}[t]
\centering
\includegraphics[height=5.45cm]{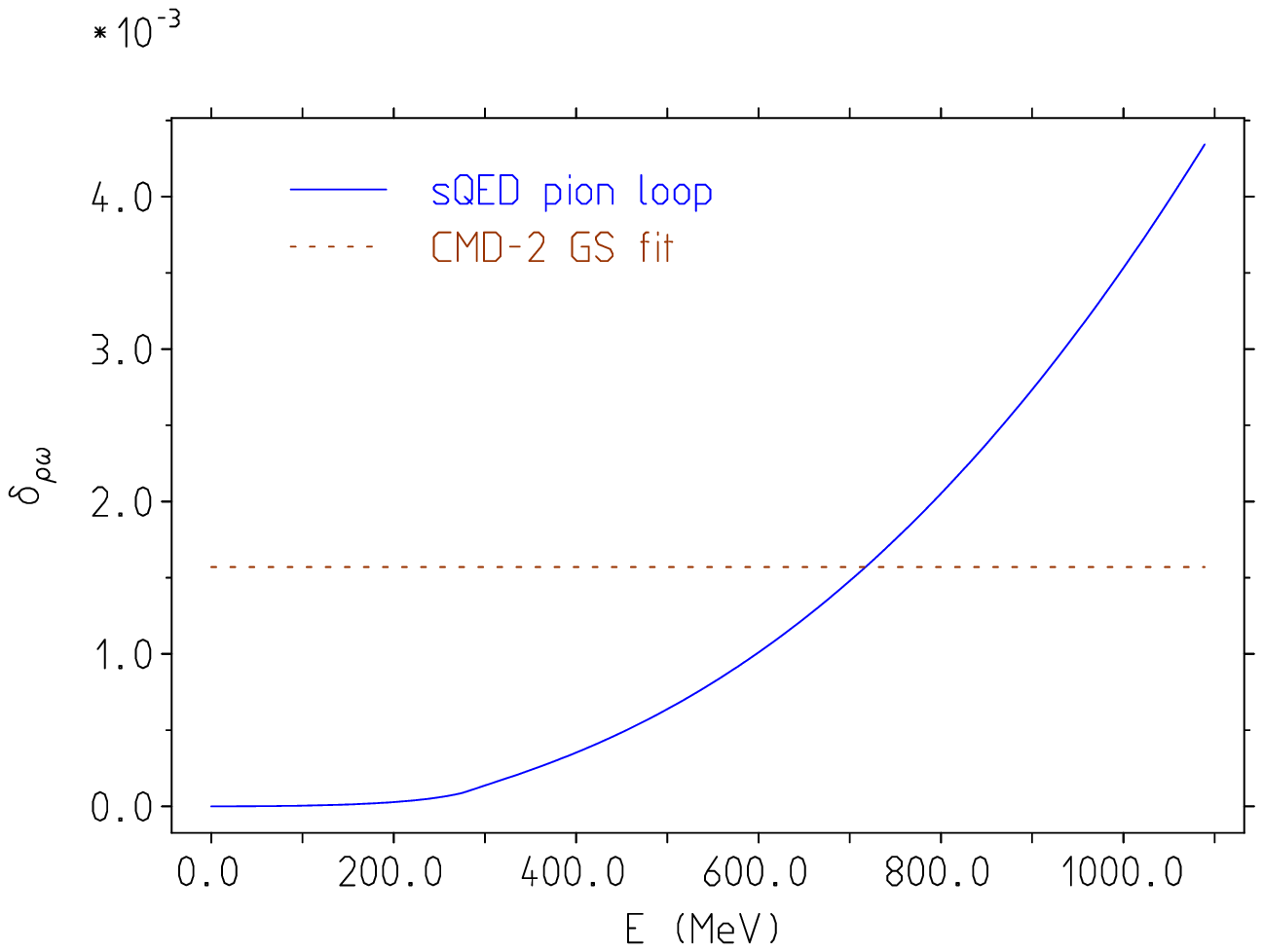}
\includegraphics[height=5cm]{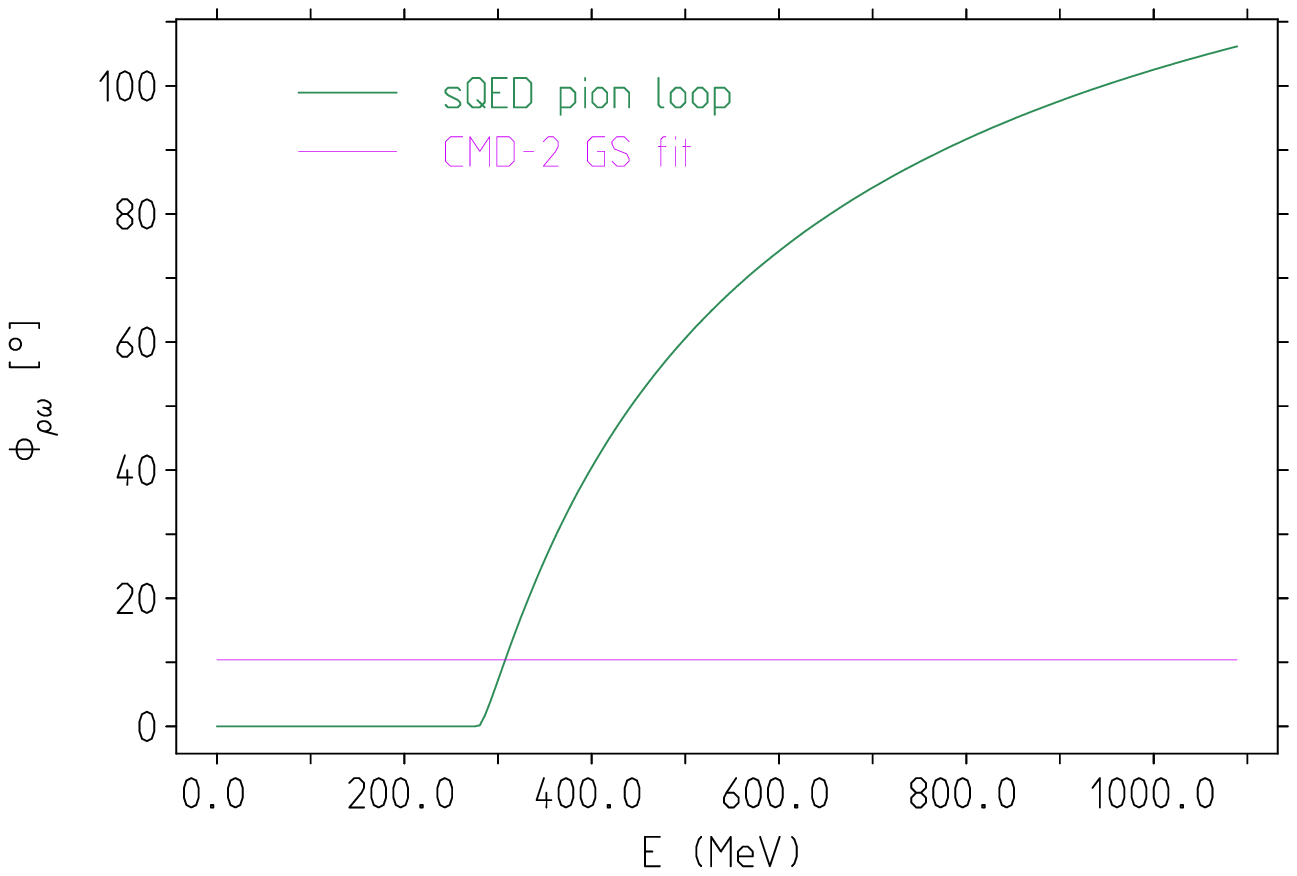}
\caption{Dynamical mixing parameter $\delta(E)$ obtained in our EFT, in contrast to the
approximation by a constant. The latter seems justified by the narrow
width of the $\omega$.}
\label{fig:}
\end{figure}

So far we have extended our effective Lagrangian by including direct
$\rho-\omega$, $\gamma-\omega$, $\omega \pi\pi$ and $\omega ee$
vertices only, such that at the one-loop level only the previous pion
loops show up. Missing are $\omega \pi^+\pi^-\pi^0$ and $\omega \pi^0
\gamma$ effective vertices, which are necessary in order to obtain
the correct full $\omega$-width in place of the
$\omega\to\pi\pi$ partial width only. Since the $\omega$ is very
narrow we expect to obtain a good approximation if we use the proper
full width in $\Impa \Pi_{\omega \omega}
=\I\,M_\omega\Gamma_\omega(s)$, namely,
\ba
 \Gamma_\omega \to \Gamma_\omega (s) &=& \sum\limits_X\,\Gamma (\omega \ra X,s) =
\frac{s}{M_\omega^2}\Gamma_\omega \left\{\sum\limits_X\,
Br(\omega \ra X) \frac{F_{X}(s)}{F_{X}(M^2_\omega)}
\right\},
\ea
where $Br(V \ra X )$ denotes the branching fraction for the channel
$X=3\pi,\pi^0 \gamma,2\pi$ and $F_{X}(s)$ is the phase space function
for the corresponding channel normalized such that $F_{X}(s) \ra$
const for $s \ra \infty$~\cite{Achasov76}.

If we include $\omega-\rho$ mixing in the usual way (see (\ref{GSform})) by writing
\ba
F_\pi(s) =  \left[e^2\,D_{\gamma\gamma}+ e\,(g_{\rho\pi\pi}-g_{\rho ee})\,D_{\gamma \rho}-
g_{\rho ee} g_{\rho\pi\pi} \,D_{\rho\rho}\cdot \left(1+\delta \frac{s}{\mr}BW_\omega(s)\right)\right]/\left[e^2\,D_{\gamma\gamma}\right]\epo
\label{fpidecompwithomega}
\ea
with $\mathrm{BW}_\omega(s)=-\mo/((s-\mo)+\I\,M_\omega \Gamma_\omega(s))$ in our
approach $\delta_{\mathrm{eff}}(s)$ is given by
\ba
\delta_{\mathrm{eff}}(s)=
\frac{(g_{\rho ee} g_{\om\pi\pi}+g_{\om ee}g_{\rho\pi\pi})\,D_{\rho \omega}
-e\,(g_{\om\pi\pi}-g_{\om ee})\,D_{\gamma \rho}}
{(g_{\rho\pi\pi}g_{\rho ee})\,D_{\rho \rho}\cdot \mathrm{BW}_\omega(s)}
\ea
which is well approximated by
\ba
\delta_{\mathrm{dyn}}&=&-\frac{(g_{\rho e e} g_{\omega \pi\pi}+g_{\omega
ee}g_{\rho \pi\pi})}{g_{\rho \pi\pi}g_{\rho ee}}\,
\frac{\Pi^{\mathrm{ren}}_{\rho\omega}(s)}{\mo} \crn &\stackrel{s\sim\mo}{\sim}&
-\frac{(g_{\rho e e} g_{\omega \pi\pi}+g_{\omega
ee}g_{\rho \pi\pi})}{g_{\rho \pi\pi}g_{\rho ee}}\,
\frac{g_{\rho\pi\pi}g_{\om\pi\pi}}{48 \pi^2}\,\left(h(\mo)-\Repa h(\mr)\right)\epo
\ea
The second term $g_{\om ee}g_{\rho \pi\pi}\sim 0.03$ is an order of
magnitude larger than than the first one $g_{\rho e e} g_{\omega
\pi\pi}\sim 0.003$ and thus is sensitive to $g_{\om ee}$ once the
$g_{\rho \pi\pi}$ has been fixed in the $\rho$-sector. In leading
approximation $\delta \propto g_{\om ee}/g_{\rho ee}\cdot g_{\rho
\pi\pi}g_{\om \pi\pi}$. The phase is actually fixed by the pion loop
alone as we take couplings to be real (unitarity). We have
$|\delta|=1.945\power{-3}$ and $\phi_\delta=90.49^\circ$.

A complete EFT treatment of the $\rho-\omega$ mixing, as well as the
proper inclusion of the higher $\rho$'s, requires the extension of our
model, e.g. in the HLS version as performed
in~\cite{Benayoun07,Benayoun10}. This
is beyond the scope of the present study. Nevertheless, the discussion
of the $\rho-\omega$ mixing presented above illustrates the need for a
reconsideration of the subject.

\section{Summary and Conclusions}
Our main point is to properly take into account the $\rho-\gamma$
mixing, which is responsible for the major part of the $\tau$
vs. $\epm$ discrepancy. The general message we have is that only a
consequent application of the effective field theory approach can help
to make progress in understanding low energy hadron data. Such
attempts have been made recently within the HLS effective
theory~\cite{Benayoun07,Benayoun10}, where also a common consistent
description (global HLS-model fit) of $\epm$- and $\tau$-data has been found
within a $(\rho,\omega,\phi)$ mixing scheme. We have some difficulties
to understand details of these elaborate calculations, but Fig.~8 of
of~\cite{Benayoun07} includes a component which has to be attributed
to the $\rho-\gamma$ mixing, what we have been discussing here (our
Fig.~\ref{fig:mixingcorr}). Of course the effects considered here can
easily be incorporated in other approaches like the ones proposed
in~\cite{Guerrero:1997ku} or~\cite{LeCo02}.

We have based our ``modeling'' of the pion from factor on the low
energy effective field theory of $\rho$, $\pi\pi$ and $\gamma$, with
the main assumption the pions to behave as point particles (sQED).  We
avoid some ad hoc elements, like imposing $F_\pi(0)=1$ by hand, which
is common practice when using GS like ans\"atze. Our result
demonstrates two things: a) models should in any case be based on
effective field theory, the ``right'' Lagrangian, to be sorted out by
global fit strategies,
b) obviously our simplest model has to be extended towards a full
fledged resonance Lagrangians in order to be able to control the
$\rho-\omega$ mixing and the energy range above the two kaon
threshold. A proper treatment of the $\rho-\omega$ mixing requires an
extension of the model to include the $\omega 3\pi$ and
$\omega\pi\gamma$ couplings, which have anomalous parity, like $\pi^0
\gamma \gamma$. Our study should be understood as a
first step towards reconsidering the proper extraction of resonance
parameters from experiments.

In spite of the fact that we have to make use of a model which has its
limitations, the relevant $\rho-\gamma$ mixing effect (needed for
example to correct the $\tau$ data to be applicable for the evaluation
of $\amuh$) can be determined from the $\epm$-data solely. Hence, the
$\tau$ data represent independent additional information, which can be
used to improve evaluations of hadronic effects which initially are
directly related to $\epm$-data.
 
With this additional insight the original idea promoted
in~\cite{ADH98} indeed can work at the level of present standards in
precision. Nevertheless, we should keep in mind that photon radiation
effects from the composite hadrons are not fully under control and
corresponding uncertainties are not easy to specify beyond the few per
mil level.

The model we use reproduces in some approximation the standard
Gounaris-Sakurai model and improves it in several respects: as it
should be we need no photon mass renormalization (which is
intrinsically there in the GS model, as explained above), in addition
to the $s$-dependence of the width (as incorporated properly in most
versions of the GS model) we include the $s$-dependent $\rho - \gamma$
mixing, which leads to substantial modification of the GS model. The
current conservation condition $F_\pi(0)=1$ is realized in our
approach in a natural way (just by gauge invariance and by standard
electromagnetic charge renormalization) as for the gauge invariant VMD
type (\ref{Lrhogamma}) and not by hand as it is done in the GS model,
which is of the VMD type (\ref{VMDnaive}).

What does it mean for the muon $g-2$? It definitely shows that the
$\tau$-decay isovector form factor must be corrected also for $\rho
-\gamma $ mixing interferences, which means that relevant corrections
not accounted for so far must be applied. These corrections are model
dependent to some extent, as we assume pions to be point-like. Our
calculation shows that the bulk of the effect is real and a 10\%
uncertainty seems to be a reasonable guess. The effects which only
depend on the $\rho$ parameters mass, width and leptonic branching
fraction and for reasonable values of these parameters brings into
fair agreement $\tau$-data based evaluations of $\amuh$ and the
$\epm$-based ones. Thus phenomenologically, it reproduces rather
precisely the pattern of the discrepancy between $\tau$ and $\epm$
extracted pion form factors (modulo differences which show up in the
different measurements anyway).  Our result strongly supports that the
observed muon $g-2$ discrepancy between theory and experiment is real
and at the 3 $\sigma$ level (see
e.g.~\cite{Jegerlehner:2009ry},\cite{Hoecker:2010qn}). The $\tau$-data
if properly corrected for isospin violating effects support this
conclusion.

For the lowest order hadronic vacuum polarization (VP) contribution to
$\amu$ we find
$$a_\mu^{\mathrm{had,LO}}[e,\tau]=690.96(1.06)(4.63)\power{-10}~~~~
(e+\tau)$$ for the higher order vacuum polarization terms we find
$-206.68(0.36)(1.56)\power{-11}$, $103.89(0.16)(0.70)\power{-11}$ and
$3.0(0.1)\power{-11}$, which adds to
$$a_\mu^{\mathrm{had,VP,HO}}=-99.79(0.38)(0.86)\power{-11}$$
(systematic errors of the first two essentially anti-correlated). The
corresponding updated value for the muon $g-2$ is
$\amu^{\mathrm{the}}=116591797(60)\power{-11}$ which deviates from the
experimental value
$\amu^{\mathrm{exp}}=116592080(54)(33)\power{-11}$~\cite{BNL04} by
$\amu^{\mathrm{exp}}-\amu^{\mathrm{the}}=(283 \pm 87)\power{-11}$
corresponding to $3.3\,\sigma$.

It is important to note that the $\rho-\gamma$ mixing effect discussed
here is included when evaluating the electromagnetic current
correlator, $F_\pi(s)$ and $\amuh$ from first principles in lattice
QCD~\cite{Renner:2010zj,Brandt:2010ed}. By comparison with the charged
channel isovector correlator, the ratio $F_0(s)/F_-(s)$ could be
measured within QCD, without reference to sQED or other hadronic
models.\\[5mm]

\noindent
{\bf Acknowledgments} \\ One of us (F.J.) gratefully acknowledges
support by the Foundation for Polish Science at an early stage of this
work. This work was supported in part by the TARI program under
contract RII3-CT-2004-506078. R.S. acknowledges a scholarship from the 
UPGOW project co-financed by the European Social Fund. We thank G. Venanzoni
for helpful remarks and suggestions.

\end{document}